\documentclass[12pt,a4paper]{report}
\usepackage{listings}
\usepackage{xcolor}
\definecolor{verylightgray}{rgb}{.97,.97,.97}

\lstdefinelanguage{Solidity}{
	keywords=[1]{anonymous, assembly, assert, balance, break, call, callcode, case, catch, class, constant, continue, constructor, contract, debugger, default, delegatecall, delete, do, else, emit, event, experimental, export, external, false, finally, for, function, gas, if, implements, import, in, indexed, instanceof, interface, internal, is, length, library, log0, log1, log2, log3, log4, memory, modifier, new, payable, pragma, private, protected, public, pure, push, require, return, returns, revert, selfdestruct, send, solidity, storage, struct, suicide, super, switch, then, this, throw, transfer, true, try, typeof, using, value, view, while, with, addmod, ecrecover, keccak256, mulmod, ripemd160, sha256, sha3}, 
	keywordstyle=[1]\color{blue}\bfseries,
	keywords=[2]{address, bool, byte, bytes, bytes1, bytes2, bytes3, bytes4, bytes5, bytes6, bytes7, bytes8, bytes9, bytes10, bytes11, bytes12, bytes13, bytes14, bytes15, bytes16, bytes17, bytes18, bytes19, bytes20, bytes21, bytes22, bytes23, bytes24, bytes25, bytes26, bytes27, bytes28, bytes29, bytes30, bytes31, bytes32, enum, int, int8, int16, int24, int32, int40, int48, int56, int64, int72, int80, int88, int96, int104, int112, int120, int128, int136, int144, int152, int160, int168, int176, int184, int192, int200, int208, int216, int224, int232, int240, int248, int256, mapping, string, uint, uint8, uint16, uint24, uint32, uint40, uint48, uint56, uint64, uint72, uint80, uint88, uint96, uint104, uint112, uint120, uint128, uint136, uint144, uint152, uint160, uint168, uint176, uint184, uint192, uint200, uint208, uint216, uint224, uint232, uint240, uint248, uint256, var, void, ether, finney, szabo, wei, days, hours, minutes, seconds, weeks, years},	
	keywordstyle=[2]\color{teal}\bfseries,
	keywords=[3]{block, blockhash, coinbase, difficulty, gaslimit, number, timestamp, msg, data, gas, sender, sig, value, now, tx, gasprice, origin},	
	keywordstyle=[3]\color{violet}\bfseries,
	identifierstyle=\color{black},
	sensitive=false,
	comment=[l]{//},
	morecomment=[s]{/*}{*/},
	commentstyle=\color{gray}\ttfamily,
	stringstyle=\color{red}\ttfamily,
	morestring=[b]',
	morestring=[b]"
}

\lstdefinelanguage{Cairo}{
	keywords=[1]{assert, func, let, tempvar, return, end, import, int}, 
	keywordstyle=[1]\color{blue}\bfseries,
	keywords=[2]{felt, HashBuiltin},	
	keywordstyle=[2]\color{teal}\bfseries,
	keywords=[3]{@l1_handler, ap, fp, memory},	
	keywordstyle=[3]\color{violet}\bfseries,
	identifierstyle=\color{black},
	sensitive=false,
	comment=[l]{\#},
	morecomment=[s]{/*}{*/},
	commentstyle=\color{gray}\ttfamily,
	stringstyle=\color{red}\ttfamily,
	morestring=[b]',
	morestring=[b]"
}

\usepackage{forest}
\usepackage{amssymb}
\usepackage{biblatex} 
\usepackage{pgfplots}
\usepackage{caption}
\usepackage{csquotes}
\usepackage{filecontents}
\usepackage{tikz}
\usepackage{booktabs}
\usepackage{multirow}
\usepackage{lipsum} 
\usepackage{mathtools}
\usepackage{newlfont}
\usepackage[framemethod=TikZ]{mdframed}
\usetikzlibrary{calc,shapes,arrows,angles,positioning,intersections,quotes,decorations.markings}
\usepackage{tkz-euclide}
\pgfplotsset{compat=newest}
\usetikzlibrary{decorations.pathmorphing}
\usetikzlibrary{pgfplots.groupplots}
\usepgfplotslibrary{dateplot}
\usepackage[english]{babel}
\usepackage{float}
\usepackage{amsmath}
\usepackage[]{svg}
\usepackage{amsfonts}
\usepackage[frozencache,cachedir=.]{minted}
\begin{document}
\newcounter{theo}[section]
\setcounter{theo}{0}
\renewcommand{\thetheo}{\arabic{theo}}
\newenvironment{theo}[2][]{%
    \refstepcounter{theo}
    \ifstrempty{#1}%
{\mdfsetup{%
    frametitle={%
        \tikz[baseline=(current bounding box.east),outer sep=0pt]
        \node[anchor=east,rectangle,fill=blue!20]
        {\strut Example~\thetheo};}
    }%
}{\mdfsetup{%
    frametitle={%
        \tikz[baseline=(current bounding box.east),outer sep=0pt]
        \node[anchor=east,rectangle,fill=blue!20]
        {\strut Example~\thetheo:~#1};}%
    }%
}%
\mdfsetup{%
    innertopmargin=10pt,linecolor=blue!20,%
    linewidth=2pt,topline=true,%
    frametitleaboveskip=\dimexpr-\ht\strutbox\relax%
}
 
\begin{mdframed}[]\relax}{%
\end{mdframed}}
\begin{titlepage}
\begin{center}
{{\Large{\textsc{Alma Mater Studiorum $\cdot$ University of Bologna}}}} \rule[0.1cm]{15.8cm}{0.1mm}
\rule[0.5cm]{15.8cm}{0.6mm}
{\small{\bf COMPUTER SCIENCE AND ENGINEERING DEPARTMENT\\
Bachelor in Computer Science }}
\end{center}
\vspace{15mm}
\begin{center}
{\LARGE{\bf Optimistic and Validity Rollups:}}\\
\vspace{3mm}
{\LARGE{\bf Analysis and Comparison between}}\\
\vspace{3mm}
{\LARGE{\bf Optimism and StarkNet}}\\
\end{center}
\par
\noindent
\vspace{20mm}
\begin{center}
{\large{\bf Luca Donno}}
\end{center}
\begin{center}
{luca.donno@studio.unibo.it}
\end{center}
\end{titlepage}

\tableofcontents

\chapter{Introduction}


A blockchain is a distributed data structure formed by a list of concatenated blocks \footnote{rarely this structure is represented as a directed acyclic graph, as in the Nano (XNO) blockchain.}. A block consists of an ordered sequence of transactions used in a state transition and a header that contains information such as the block number, the hash of the previous block header, and the root of the Merkle tree representing the transactions. 

A Merkle tree \cite{merkle1987digital} is a data structure where each leaf consists of the hash of a block of data and each internal node is computed as the hash of its children. In this way, the root succinctly represents the underlying data, and the inclusion of a leaf can be proved with a number of steps and a proof size proportional to the height of the tree, i.e., $O(\log n)$. Ethereum uses a variant of the Merkle Patricia tree \cite{morrison1968patricia} to optimize its proofs and searches, but for the purposes of this writing it is sufficient to think of them as Merkle trees.

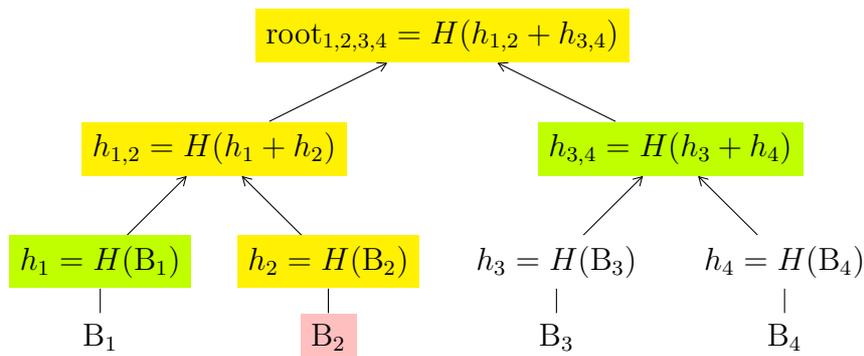
\begin{figure}[H]
\centering
\tikzstyle{level 1}=[level distance=15mm, sibling distance=60mm]
\tikzstyle{level 2}=[level distance=15mm, sibling distance=30mm]
\tikzstyle{level 3}=[level distance=10mm]
\begin{tikzpicture}[grow=down,<-,>=angle 60]
\begin{scope}[yshift=0]
  \node[fill=yellow] {$\text{root}_{1,2,3,4} = H(h_{1,2}+h_{3,4})$}
    child {node[fill=yellow] {$h_{1,2} = H(h_1+h_2)$}
      child {node[fill=lime] {$h_1 = H(\text{B}_1)$}
        child[-] {node{$\text{B}_1$}}  
      }
      child {node[fill=yellow]{$h_2 = H(\text{B}_2)$}
        child[-] {node[fill=pink]{$\text{B}_2$}}  
      }
    }
    child {node[fill=lime] {$h_{3,4} = H(h_3+h_4)$}
      child {node{$h_3 = H(\text{B}_3)$}
        child[-] {node{$\text{B}_3$}}  
      }
      child {node{$h_4 = H(\text{B}_4)$}
        child[-] {node{$\text{B}_4$}}  
      }
    };
\end{scope}
\end{tikzpicture}
\caption{A Merkle tree. To prove inclusion in the root of block $B_2$ it is sufficient to provide a logarithmic number of nodes (in green) with respect to the number of blocks, while for verification it is sufficient to calculate a logarithmic number of hashes (in yellow).}
\end{figure}

In simpler blockchains, such as Bitcoin, transitions consist of updating the native token allocations to users, while in blockchains that integrate a virtual machine, such as Ethereum \footnote{with the Ethereum Virtual Machine (EVM).}, they can interact with arbitrary programs (called \textit{smart contract}) and modify its memory (called \textit{storage}).

Each node in the network contains a replica of the current state of the blockchain to which blocks are added at regular intervals. Despite their decentralized structure, they are logically centralized in that each node must observe the same version of facts. 
Since everyone can add a block at the top of the blockchain, multiple versions of the new list are proposed to the network and each of these is named \textit{fork}.
For this reason, blockchains implement a consensus algorithm to decide which of the possible versions of the state to follow and to which new transactions to add, called the \textit{fork choice rule}. The two most frequently used categories of consensus algorithms are \textit{Proof of Work} and \textit{Proof of Stake}.

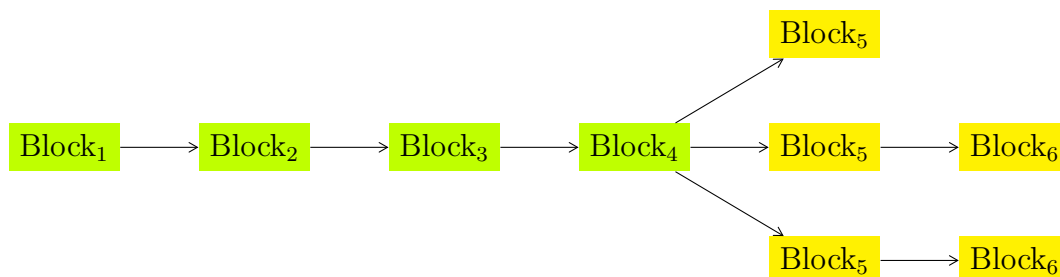
\begin{figure}[H]
\centering
\tikzstyle{level 1}=[level distance=25mm, sibling distance=30mm]
\tikzstyle{level 2}=[level distance=25mm, sibling distance=15mm]
\tikzstyle{level 3}=[level distance=25mm]
\begin{tikzpicture}[grow=right,->,>=angle 60]
\begin{scope}[yshift=0]
  \node[fill=lime] {$\text{Block}_1$}
    child {node[fill=lime] {$\text{Block}_2$}
      child{ node[fill=lime] {$\text{Block}_3$}
        child {node[fill=lime] {$\text{Block}_4$}
            child {node[fill=yellow] {$\text{Block}_5$}
                child {node[fill=yellow] {$\text{Block}_6$}}
            }
            child {node[fill=yellow] {$\text{Block}_5$}
                child {node[fill=yellow] {$\text{Block}_6$}}
            }
            child {node[fill=yellow] {$\text{Block}_5$}
            }
        }  
      }
    };
\end{scope}
\end{tikzpicture}
\caption{A blockchain with three forks from block 4. In green are the blocks on which the network has reached consensus, in yellow the others.}
\end{figure}

Nodes, in addition to following the consensus rule, verify that transactions in a block comply with validity rules: for example, a user cannot spend more coins than they own, a user cannot make transactions using another user's account, a block cannot claim that the result of $2+2$ is not $4$.

A node in the network can independently verify these rules, which have priority over consensus rules: if a block is chosen by the consensus rule but contains invalid transactions, it is discarded.
Nodes that verify both the validity of blocks and the consensus rule are named \textit{full node}. Using one's own full node to interact with a blockchain is critical so as not to rely on any intermediary, which might show a malicious version of its state.

The throughput of a blockchain, which is the number of transactions per second (TPS), is proportional to the size of the blocks, which determines how many transactions a block can hold, and their frequency over time. Since the space in a block is finite, the cost to include a transaction is a function of the supply and demand for it, which is why a higher throughput is desired.

\newpage

The size of a block can be defined in bytes or in \textit{gas}. Gas represents a unit of measurement of the cost, in terms of resources used, of a computation. For example, a simple Ether transfer on Ethereum costs $21000$ gas.

\begin{table}[H]
  \begin{center}
    \label{tab:table1}
    \begin{tabular}{l|c|r} 
      \textbf{Opcode} & \textbf{Gas}\\
      \hline
      ADD & 3\\
      XOR & 3\\
      MUL & 5\\
      MULMOD & 8\\
      JUMP & 8\\
      BLOCKHASH & 20
    \end{tabular}
    \caption{Some EVM opcodes and their respective costs in gas.}
  \end{center}
\end{table}

Different blockchains have different performances:

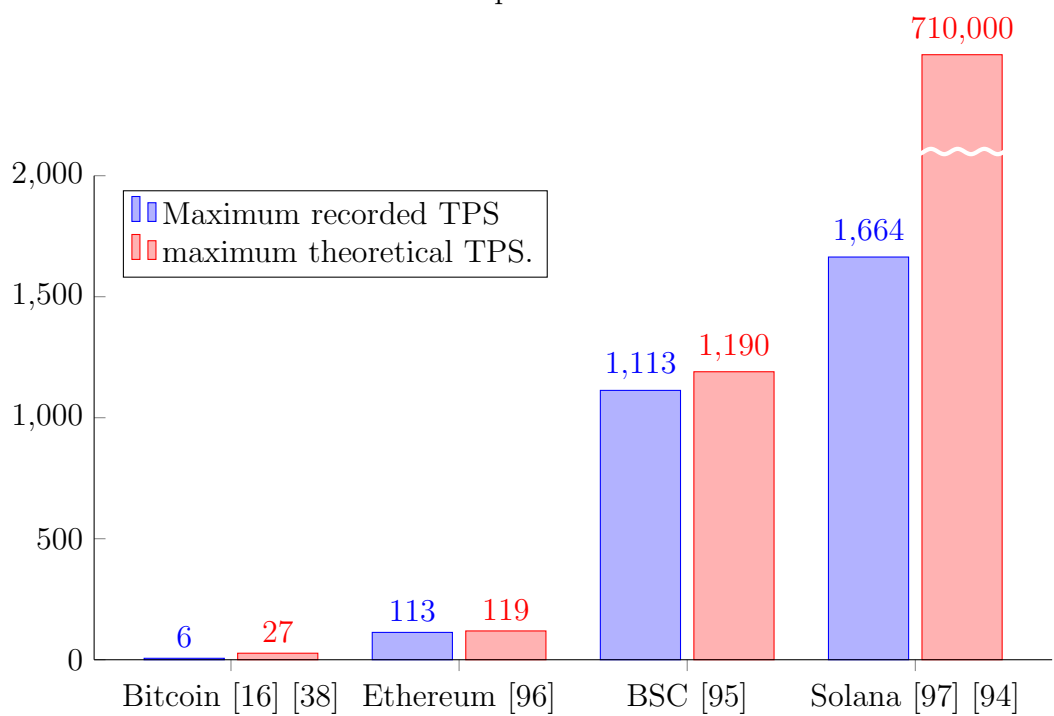
\begin{tikzpicture}
\begin{axis}[
    every axis plot post/.style={/pgf/number format/fixed},
    legend pos=north west,
    legend cell align={left},
    ybar=5pt,
    height=8cm,
    bar width=30pt,
    x=3cm,
    ymin=0,
    axis on top,
    ymax=2000,
    xtick=data,
    enlarge x limits=0.2,
    symbolic x coords={Bitcoin \cite{bitcointps2022} \cite{georgiadis2019many},Ethereum \cite{ethtps2022},BSC \cite{bsctps2022},Solana \cite{solanatps2022} \cite{yakovenko2018solana}},
    restrict y to domain*=0:2500, 
    visualization depends on=rawy\as\rawy, 
    after end axis/.code={ 
            \draw [ultra thick, white, decoration={snake, amplitude=1pt}, decorate] (rel axis cs:0,1.05) -- (rel axis cs:1,1.05);
        },
    nodes near coords={%
            \pgfmathprintnumber{\rawy}
        },
    axis lines*=left,
    clip=false
    ]
\addplot coordinates {(Bitcoin \cite{bitcointps2022} \cite{georgiadis2019many},6) (Ethereum \cite{ethtps2022},113) (BSC \cite{bsctps2022},1113) (Solana \cite{solanatps2022} \cite{yakovenko2018solana}, 1664)};
\addplot coordinates {(Bitcoin \cite{bitcointps2022} \cite{georgiadis2019many},27) (Ethereum \cite{ethtps2022},119) (BSC \cite{bsctps2022},1190) (Solana \cite{solanatps2022} \cite{yakovenko2018solana}, 710000)};
\legend {Maximum recorded TPS, maximum theoretical TPS.};
\end{axis}
\end{tikzpicture}

\section{The gas crisis}

Because of the high utilization of the network \cite{ethutilization2022} and limited throughput, Ethereum has shown how the gas price can increase by two orders of magnitude in a single day \cite{ethereumgasprice2022} due to single events. For example, during the release of the UNI token of the Uniswap protocol \cite{unilaunch2020}, the gas reached an average price of about $538$ Gwei \footnote{1 Gwei equals $10^9$ wei. $10^{18}$ wei $= 1$ ETH, so one billion Gwei equals 1 Ether.}, with a maximum of $133053$ Gwei. A simple token exchange at 488 Gwei \cite{tokenswap2020uni} costs about 0.055 ETH in fees, which at the price of 1800\$ is equivalent to 99\$. More recently, on May 1, 2022, during the release of the NFT Otherside \cite{otherside} collection, some users spent more than \$4,000 in commissions to obtain a pair of NFTs \cite{otherside2022mint}, bringing the average commission per transaction to \$200, the highest ever \cite{avgfee2022}. Vitalik Buterin, the creator of Ethereum, stated in a famous 2017 interview \cite{vitalik5cents2022} that \enquote{the Internet of money should not cost more than 5 cents per transaction}.

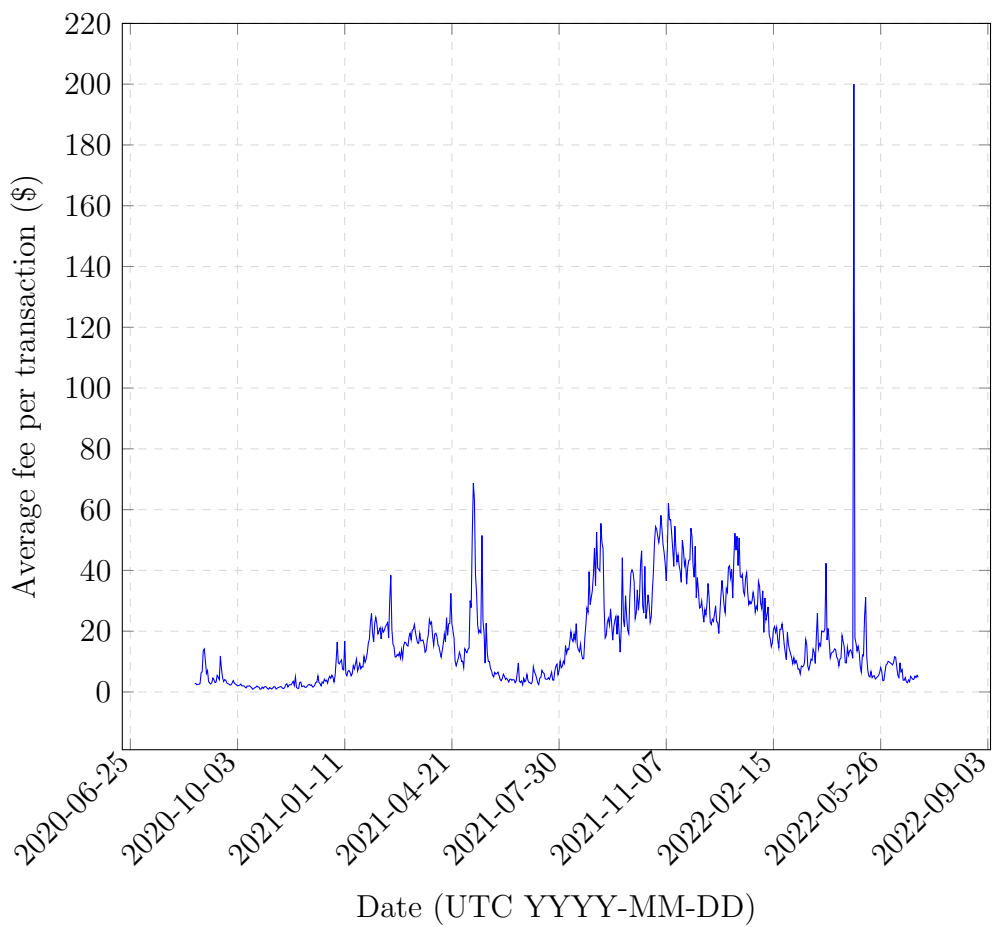
\begin{figure}[H]
\centering
\begin{tikzpicture}
      \begin{axis}[
          width=\linewidth, 
          grid=major,
          width=13cm,
          grid style={dashed,gray!30},
          xlabel=Date (UTC YYYY-MM-DD), 
          ylabel=Average fee per transaction (\$),
          date coordinates in=x,
          date ZERO=2021-07-30,
          legend style={at={(0.5,-0.2)},anchor=north},
          x tick label style={rotate=45,anchor=east}
        ]
        \addplot[mark=none, blue] 
        table[x=DateTime,y=Average Txn Fee (USD),col sep=comma] {avgfee.csv}; 
        \legend{}
      \end{axis}
    \end{tikzpicture}
\end{figure}

\section{Decentralized scalability}

To increase the throughput of a blockchain, a trivial solution is to increase the size of blocks (or their frequency). In the context of Ethereum, this means increasing the maximum amount of gas a block can hold. With this method, as each full node must process every transaction of every block to verify its validity, as the throughput increases, the hardware requirements also increase, consequently leading to a greater centralization of the network. Some blockchains, such as Bitcoin and Ethereum, optimize their design to maximize their architectural decentralization, others, such as the Binance Smart Chain (BSC) and Solana, to be as fast and cheap as possible. Decentralized networks artificially limit the throughput of the blockchain to decrease the hardware requirements to participate in the network. This trade-off is called the Scalability Trilemma \cite{monte2020scaling} \cite{hafid2020scaling}: 

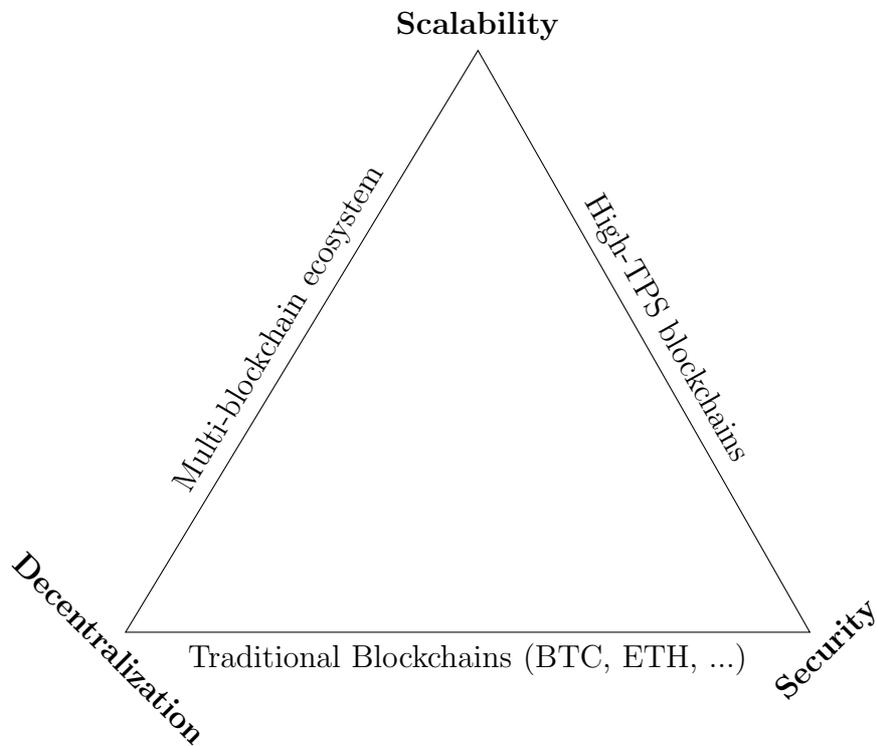
\vspace{1cm}

\begin{figure}[H]
\centering
\begin{tikzpicture}
\path (59:9) coordinate (A)
      (0:9)  coordinate (B)
      (0:0)  coordinate (C);
\draw (A)
-- (B) node [at start, above]{\textbf{Scalability}} node [midway,above, sloped]{High-TPS blockchains}
-- (C) node [at start, below, rotate=45]{\textbf{Security}} node [midway,below, sloped] {Traditional Blockchains (BTC, ETH, ...)}
-- (A) node [at start, below, rotate=-45]{\textbf{Decentralization}} node [midway, above, sloped] {Multi-blockchain ecosystem} -- cycle;
\end{tikzpicture}
\end{figure}

The Trilemma states that there is no simple technique to achieve all three properties, but only two of them. These properties are:
\begin{itemize}
    \item \textbf{Scalability}: the network processes more transactions than a simple node (e.g., an ordinary laptop) can verify.
    \item \textbf{Decentralization}: the network operates without the reliance of a small number of large centralized operators.
    \item \textbf{Security}: the network is able to withstand a large percentage of hostile nodes.
\end{itemize}

Traditional blockchains aim to have tens or hundreds of millions of independent nodes at the expense of scalability, while high-performance blockchains usually do not exceed a hundred nodes due to high hardware requirements. Solana requires at least a 16-core CPU and 256GB of RAM for an RPC node \cite{solanahwreq2022}. 

A multi-chain ecosystem consists of having different applications on different blockchains communicating with each other. This system is decentralized and scalable, but allows an attacker to take control of only one of them to cause ripple effects on other networks.

\vspace{1pt}

\section{Scalability solutions}

Over the years, attempts have been made to find a solution to solve the Trilemma. All of these solutions have the characteristic of moving some activity off-chain, linking on-chain activity to off-chain activity using smart contracts, and verifying on-chain what is happening off-chain. The three main scalability solutions are state channels, Plasma, and Rollups.

\subsection{State channels}

State channels \cite{negka2021blockchain} are related to Bitcoin payment channels \cite{mccorry2016towards} \cite{avarikioti2018towards} in that they allow instant payments without fees, but for transactions that alter the state of the virtual machine. The components of a state channel are:

\begin{enumerate}
    \item A smart contract that defines and locks the initial state to interact with.
    \item Participants who update the state among themselves by constructing and signing transactions that could be published on-chain but are not initially. Each state alteration takes the place of the previous one.
    \item A mechanism for closing the channel that unlocks the state.
\end{enumerate}

The key point is that step 2 does not need to interact with the blockchain at all, but each participant must be guaranteed to be able to update the state at any time. The limitation of this approach is that participants, or someone for them, must remain online until the channel is closed. This is not a problem for those applications that already require being online all the time, such as turn-based games.

\begin{theo}[Tic-Tac-Toe]{thm:tris}
The smart contract is initialized with an initially empty 3x3 Tic-tac-toe field and the address of the two opponents. Alice starts the game by choosing where to place her mark, signs the message containing the updated matrix, and sends it to Bob, who also signs it. The message is not posted on-chain: it is the duty of the participants, or someone for them, to keep track of it. The turn goes to Bob, who chooses the location of his own symbol, updates the matrix, signs it, and sends it to Alice. 

However, Alice can refuse to sign the message if she does not like Bob's choice: therefore, there must be a mechanism to avoid this situation. Bob publishes the previous state on the smart contract, which checks its validity and signatures, and executes his current on-chain move. If Alice refuses to continue playing (or loses the connection), a time-out can be implemented that eventually awards the win to Bob. 

If, on the other hand, both players are willing to play until the end of the game, the only state that is published on the smart contract is the final state, where a function verifies its configuration and awards the prize to the winner. 
\end{theo}

State channels are not general because the implementation depends on the specific application. They are also not suitable for those situations where the objects involved or the actions to be performed do not have an explicit participant associated with them.

\subsection{Plasma}

A Plasma chain \cite{poon2017plasma} is an autonomous blockchain that is anchored to Ethereum and performs off-chain transactions. It is managed through a smart contract on the main blockchain that handles deposits and withdrawals. The state of Plasma is published periodically on Ethereum in the form of a Merkle root. To gain access, users deposit an amount of tokens (ETH or any ERC-20 token \cite{vogelsteller2015eip20}), which are recreated in equal amounts on the Plasma chain.

To withdraw funds on Ethereum, the mechanism is more complicated, as Ethereum is unable to verify, having only the Merkle root, that the Plasma state is correct. A malicious user could create a fictitious state claiming that he owns 1000 ETH, and publish a Merkle proof of inclusion in the Merkle root. For this reason, a dispute period is implemented, usually 7 days, in which anyone can prove that a Merkle proof is invalid, invalidating the withdrawal. 

Correct behavior can be incentivized by having an amount of ETH deposited before withdrawal, which is subtracted in the case of invalid behavior.

The two most widely used implementations of Plasma are Plasma MVP \cite{johnson2019sidechains} and Plasma Cash \cite{konstantopoulos2019plasma}.

\begin{theo}[Double spending]{thm:doublespending}
    Alice uses Plasma MVP to perform its transactions, using a UTXO \footnote{Unspent Transaction Output \cite{delgado2018analysis}} model. Alice sends tokens to Bob creating a transaction using a UTXO. Alice later wants to use the same UTXO, which is now Bob's, to withdraw funds.
    Bob shows, on the contract representing the Plasma chain on Ethereum, that Alice has already spent the UTXO in a previous transaction. If Bob is offline instead, Alice is successful in making a double spend \cite{chohan2021double}.
\end{theo}

Plasma, like state channels, is not a general scalability solution as it does not support the use of smart contracts \cite{barbie2018plasma}.

The main limitation arises from the Data Availability Problem (DAP): if an operator publishes an invalid state transition, users cannot create invalidity proofs if it does not publish data corresponding to Merkle roots.

\subsection{Rollup}

Rollups are blockchains that publish their blocks on another blockchain, called Layer 2 (L2) and Layer 1 (L1) respectively, inheriting its consensus and data availability \cite{marukhnenko2021overview}. 

Rollups have three main components:
\begin{itemize}
    \item \textbf{Sequencers}: nodes that receive Rollup transactions from users and combine them into a block that they send to Layer 1. The block consists of at least the state root (e.g., as Merkle root) and transaction data, solving the Data Availability Problem.
    \item \textbf{Rollup full nodes}: nodes that obtain blocks from Layer 1, process and validate all transaction data by verifying that the root is correct. If a block contains invalid transactions it is discarded. In this way, Sequencers cannot create valid blocks containing invalid transactions.
    \item \textbf{Rollup light nodes}: nodes that obtain blocks from Layer 1 except for transactions. They cannot compute the new state themselves, but verify that it is valid using techniques such as invalidity or validity proofs.
\end{itemize}

Rollups use the ordering of Layer 1 blocks to determine the head of the blockchain. A block is said to be finalized if it is the first valid block at its height. They, unlike other solutions, support arbitrary computation. They are scalable because the amortized cost of transactions decreases as the number of users increases, as the cost of ensuring blockchain validity grows sub-linearly relative to the cost of verifying transactions individually.

Rollups differ based on the mechanism by which they ensure the validity of transaction execution at light nodes: in Optimistic Rollups it is ensured through an economic model and invalidity proofs (sometimes called fraud or fault proofs), while in Validity Rollups it is mathematically ensured by validity proofs. 

Because transaction execution no longer takes place on the base blockchain but off-chain, people have begun to think of blockchains no longer as monolithic but modular structures. Light nodes can be implemented as smart contracts on Layer 1: they accept the root of the new state and verify validity or invalidity proofs. These Rollups are called Smart Contract Rollups; if light nodes are independent instead, they are called Sovereign Rollups. The advantage of using a Smart Contract Rollup is being able to build a bridge between the two blockchains: since the validity of the state of L2 is proven to L1, a system of transactions from L2 to L1 can be implemented, allowing withdrawals. The disadvantage is that the cost of transactions thus depends on the cost of verifying the state on L1: if the base layer is saturated by other activities, the cost of transactions on the Rollup also increases; a cost that is avoided by using a Sovereign Rollup instead. The data and consensus layers are the ones that determine the security of the system as they define the ordering of transactions, prevent attacks, and make data available to prove state validity. In some constructs called Validiums, the data can be published on a different network from the one where the consensus occurs, but in this case new security assumptions are introduced and that is why they are not considered Rollups.

\renewcommand{\arraystretch}{2}
\begin{table}[H]
\begin{tabular}{r|c|c|c|c|}
\cline{2-5}
\multicolumn{1}{l|}{}               & \textbf{Monolithic}       & \textbf{Smart Con. Rollup} & \textbf{Sovereign Rollup}   & \textbf{Validium} \\ \hline
\multicolumn{1}{|r|}{\textbf{Data}} & \multirow{4}{*}{Ethereum} & \multirow{3}{*}{Ethereum}  & \multirow{2}{*}{Ethereum} & Off-chain         \\ \cline{1-1} \cline{5-5} 
\multicolumn{1}{|r|}{\textbf{Consensus}}   &  &                   &                                 & \multirow{2}{*}{Ethereum} \\ \cline{1-1} \cline{4-4}
\multicolumn{1}{|r|}{\textbf{Settlement}}   &  &                   & \multirow{2}{*}{Sovereign Rollup} &                           \\ \cline{1-1} \cline{3-3} \cline{5-5} 
\multicolumn{1}{|r|}{\textbf{Execution}} &  & Smart Con. Rollup &                                 & Validium                  \\ \hline
\end{tabular}
\caption{\label{tab:table-name}Types of Rollups possible on Ethereum. Validiums are not considered Rollups because they add security assumptions. If it allows the user to choose whether to publish data on-chain or off-chain, the solution is named Volition \cite{smith2022volition}.}
\end{table}

\newpage

\section{Structure of the dissertation}

In the second chapter, Optimistic Rollups are introduced and in particular Optimism Bedrock is discussed, going into detail about deposits, sequencing, withdrawals, and the invalidity proof system.

In the third chapter, Validity Rollups are introduced along with the presentation of the different proof techniques studied in the field of computational cryptography: probabilistic proofs, interactive proofs, zero-knowledge proofs, SNARK and STARK proofs. Next, StarkNet is presented using a structure similar to that of Optimism Bedrock.

The fourth chapter is a comparison of the two technologies: an analysis of the rationale behind the difference in withdrawal times, the possibility of applying the technology recursively (L3 and beyond), transaction cost with ad hoc solutions to minimize it, compatibility with Ethereum, and the licenses used is carried out.

\chapter{Optimistic Rollups}

Optimistic Rollups are currently the scalability solution that locks the most value. As of Sept. 2, 2022, Arbitrum \cite{nitro2022} and Optimism \cite{mark_tyneway_2022_6894644}, the two most widely used Optimistic Rollups, collectively locks \$4.2 billion, about 80.3\% of the value locked in the totality of Rollups \cite{l2value2022}. Both of these Rollups base their design on the EVM, a choice that has allowed them to see strong development because of compatibility with existing tools for Ethereum and the use of Solidity. Fuel \cite{fuel2022} is an attempt to build an Optimistic Rollup different from the EVM: it leverages a UTXO-based model that makes it possible to process transactions in parallel and uses Sway for writing smart contracts, a language inspired by Rust.

\section{Preliminary}

The idea of accepting \textit{optimistically} the output of blocks without verifying their execution is already present in Bitcoin's whitepaper \cite{nakamoto2008bitcoin}, discussing light nodes. These nodes only follow the header chain by verifying the consensus rule, making them vulnerable to accept blocks containing invalid transactions in the event of a 51\% attack, where instead a full node would reject them. Nakamoto proposes to solve this problem by using an ``alert'' system to warn light nodes that a block contains invalid transactions. This mechanism is first implemented by Al-Bassam, Sonnino, and Buterin \cite{al2018fraud} in which an invalidity proof system based on error correction codes is used \cite{elias1954error}. In order to enable the creation of invalidity proofs, it is necessary that the data from all blocks, including invalid blocks, be available to the network: this is the Data Availability Problem, which is solved by using a probabilistic data sampling mechanism.

The first Optimistic Rollup design was presented by John Adler and Mikerah Quintyne-Collins in 2019 \cite{adler2019building}, in which blocks are published on another blockchain that defines their consensus on ordering.

\section{Optimism Bedrock}

Bedrock \cite{mark_tyneway_2022_6894644} is the latest version of Optimism. The previous version, the Optimistic Virtual Machine (OVM), required an ad hoc compiler to compile Solidity into its own bytecode: in contrast, Bedrock is fully equivalent to the EVM in that the execution engine follows the Ethereum Yellow Paper specification \cite{wood2014ethereum}.

\subsection{Overview}
The calldata of each transaction that occurs on L2 is published on L1 by Sequencers in the form of a batch.
Users can deposit transactions on Layer 2 using a smart contract on L1: this mechanism is mainly used to move assets such as tokens and NFTs.
A Rollup node can entirely reconstruct the rollup chain by reading the batches and deposited transactions on L1. Transactions are included in this blockchain only if they are valid, in exactly the same way as for Layer 1. 
In order to allow transactions to be sent from L2 to L1, Layer 1 needs to know the state of Layer 2. Merkle roots representing the state of Layer 2 and the set of withdrawals is thus saved on Layer 1. The transaction is sent on a Layer 2 contract, which is finalized by providing a Merkle proof of inclusion in the root. This can only occur at the end of a ``dispute period'', during which a chance is given to prove root invalidity using Cannon, Optimism's invalidity proof system.

\subsection{Deposits}

Users can deposit transactions through a contract on Ethereum, the Optimism Portal, by calling the \verb|depositTransaction| function.

\begin{lstlisting}[language=Solidity]
function depositTransaction(
        address _to,
        uint256 _value,
        uint64 _gasLimit,
        bool _isCreation,
        bytes memory _data
    ) public payable metered(_gasLimit) {
        // Just to be safe, make sure that people specify address(0) as the target when doing
        // contract creations.
        if (_isCreation) {
            require(
                _to == address(0),
                "OptimismPortal: must send to address(0) when creating a contract"
            );
        }

        // Transform the from-address to its alias if the caller is a contract.
        address from = msg.sender;
        if (msg.sender != tx.origin) {
            from = AddressAliasHelper.applyL1ToL2Alias(msg.sender);
        }

        bytes memory opaqueData = abi.encodePacked(
            msg.value,
            _value,
            _gasLimit,
            _isCreation,
            _data
        );

        // Emit a TransactionDeposited event so that the rollup node can derive a deposit
        // transaction for this deposit.
        emit TransactionDeposited(from, _to, DEPOSIT_VERSION, opaqueData);
    }
\end{lstlisting}

When a transaction is executed, a \verb|TransactionDeposited| event is emitted, which each node in the Rollup listens for to process deposits. A deposited transaction is a L2 transaction that is derived from L1.
If the caller of the function is a contract (verified by comparing \verb|msg.sender| and \verb|tx.origin|), the address is transformed by adding \verb|0x1111000000000000000000000000000000001111| to it. This prevents attacks in which a contract on L1 has the same address as a contract on L2, but a different code. This can occur when contracts are created using the \verb|CREATE| opcode, which calculates the contract address as \verb|keccak256(senderAddress, nonce)|.

The inclusion on L2 of deposited transactions is ensured by specification within a \textit{sequencing window}.

The deposited transactions are a new EIP-2718 compatible transaction type \cite{zoltu2020eip2718} with prefix \verb|0x7E|, where the rlp-encoded fields are:
\begin{itemize}
    \item \verb|bytes32 sourceHash|: hash that uniquely identifies the source of the transaction.
    \item \verb|address from|: the address of the sender.
    \item \verb|address to|: the receiver address, or the zero address if the deposited transaction is a contract creation.
    \item \verb|uint256 mint|: the value to be created on L2.
    \item \verb|uint256 value|: the value to be sent to the recipient.
    \item \verb|bytes data|: the input data.
    \item \verb|bytes gasLimit|: the gas limit of the transaction.
\end{itemize}

The \verb|sourceHash| is computed as the keccak256 hash of \verb|bytes32(uint256(0))|, \\ \verb|keccak256(l1BlockHash)| and \verb|bytes32(uint256(l1LogIndex))| where \verb|l1BlockHash| and \verb|l1LogIndex| uniquely identify a \verb|TransactionDeposited| event in a block.
Without a \verb|sourceHash| two different deposited transactions can have the same hash.

A deposited transaction is executed in the following steps:
\begin{itemize}
    \item \verb|from| balance is increased by \verb|mint|.
    \item \verb|CALLER| and \verb|ORIGIN| are set to \verb|from|.
    \item \verb|context.calldata| is set to \verb|data|.
    \item \verb|context.gas| is set to \verb|gasLimit|.
    \item \verb|context.value| is set to \verb|value|.
    \item the nonce of \verb|from| is incremented by 1.
\end{itemize}

\subsubsection*{Gas market}

Since deposited transactions are initiated on L1 but executed on L2, the system needs a mechanism to pay on L1 for gas spent on L2. One solution is to send ETH through the Portal, but this implies that every caller (even indirect callers) must be marked as \verb|payable|, and this is not possible for many existing projects. The alternative is to burn the corresponding gas on L1.
The gas $g$ allocated to deposited transactions is called \textit{guaranteed gas}.
L2's gas price on L1 is not automatically synchronized but is estimated using a mechanism similar to EIP-1559 \cite{buterin2019eip1559}, opening the possibility to arbitrage between the two. The maximum amount of gas guaranteed per Ethereum block is 8 million, with a target of 2 million.

The quantity $c$ of ETH required to pay for gas on L2 is $c = g * b_{\text{L2}}$ where $b_{\text{L2}}$ is the basefee on L2. The contract on L1 burns an amount of gas equal to $c / b_{\text{L1}}$.
The gas spent to call \verb|depositTransaction| is reimbursed on L2: if this amount is greater than the guaranteed gas, no gas is burned.

\subsubsection*{L1 Attributes Deposited Transaction}

The first transaction of a rollup block is a \textit{L1 attributes deposited transaction}, used to register on a L2 predeploy the attributes of Ethereum blocks.
The address of the predeploy is \verb|0x420000000000000000000000000000000000000000000000000015| and the account executing the deposits is \verb|0xdeaddeaddeaddeaddeaddead0001|, an EOA whose private key is unknown and which is returned by the opcodes \verb|CALLER| and \verb|ORIGIN| during the execution of a L1 attributes deposited transaction.

The attributes that predeploy gives access to are:
\begin{itemize}
    \item The block number of L1.
    \item The timestamp of the block of L1.
    \item The basefee of the block of L1.
    \item The hash of the block of L1.
    \item The sequence number, which is the block number of L2 relative to the associated L1 block (also called \textit{epoch}). This number is reset when a new epoch starts.
\end{itemize}

\subsection{Sequencing}

The rollup nodes derive the Optimism chain entirely from Ethereum. This chain is extended each time new transactions are published on L1, and its blocks are reorganized each time Ethereum blocks are reorganized.

The rollup blockchain is divided into epochs. For each $n$ block number of Ethereum, there is a corresponding $n$ epoch. Each epoch contains at least one block, and each block in an epoch contains a L1 attributes deposited transaction. The first block in an epoch contains all transactions deposited through the Portal. Layer 2 blocks may also contain \textit{sequenced transactions}, i.e., transactions sent directly to the Sequencer.

\subsubsection*{Sequencer}
\label{opDA}

The Sequencer accepts transactions from users and builds blocks. For each block, it constructs a \textit{batch} to be published on Ethereum. Several batches can be published in a compressed manner, taking the name \textit{channel}. A channel can be divided into several \textit{frames}, in case it is too large for a single transaction. A channel is identified by a timestamp and a random value so that they can be published in non-sequential order, as can their frames.

The fields of a frame are: \verb|channel_id|, \verb|random|, \verb|timestamp|, \verb|is_last|, \verb|frame_data|, \verb|frame_data_length|, and \verb|frame_number|.
A channel is defined as the compression with ZLIB \cite{rfc1950} of rlp-encoded batches.
The fields of a batch are: \verb|epoch_number|, \verb|epoch_hash|, \verb|parent_hash|, \verb|timestamp|, and \verb|tx_list|.

\subsubsection*{Derivation}
A sequencing window, identified by an epoch, contains a fixed number $w$ of consecutive L1 blocks that a derivation step takes as input to construct a variable number of L2 blocks. For epoch $n$, the sequencing window $n$ includes the blocks $[n, n+w)$. This implies that the ordering of L2 transactions and blocks within a sequencing window is not fixed until the window ends. A rollup transaction is called \textit{safe} if the batch containing it has been confirmed on L1.

Frames are read from L1 blocks to reconstruct batches. The current implementation does not allow the decompression of a channel to begin until all corresponding frames have been received. Invalid batches are ignored.

Individual block transactions are obtained from the batches, which are used by the execution engine to apply state transitions and obtain the rollup state.

\subsection{Withdrawals}

In order to process withdrawals, an L2-to-L1 messaging system is implemented. 

Ethereum needs to know the state of L2 in order to accept withdrawals, and this is done by publishing on the \verb|L2OutputOracle| smart contract on L1 the roots of the states of L2 for each block. These roots are optimistically accepted as valid (or finalized) if no proof of invalidity is performed during the \textit{dispute period}. Only addresses designated as \textit{Proposers} can publish output roots. The validity of output roots is incentivized by having Proposers deposit a stake that is subtracted if they are shown to have proposed an invalid root. Transactions are initiated by calling the function \verb|initiateWithdrawal| on the predeploy at \verb|0x4200000000000000000000000000000000000000| on L2 and then finalized on L1 by calling the function \verb|finalizeWithdrawalTransaction| on the previously mentioned Optimism Portal.

\begin{lstlisting}[caption={Function of the predeploy on L2 to start a withdrawal.},captionpos=b][language=Solidity]
/**
 * @notice Sends a message from L2 to L1.
 *
 * @param _target   Address to call on L1 execution.
 * @param _gasLimit Minimum gas limit for executing the message on L1.
 * @param _data     Data to forward to L1 target.
 */
function initiateWithdrawal(
    address _target,
    uint256 _gasLimit,
    bytes memory _data
) public payable {
    bytes32 withdrawalHash = Hashing.hashWithdrawal(
        Types.WithdrawalTransaction({
            nonce: nonce,
            sender: msg.sender,
            target: _target,
            value: msg.value,
            gasLimit: _gasLimit,
            data: _data
        })
    );

    sentMessages[withdrawalHash] = true;

    emit WithdrawalInitiated(nonce, msg.sender, _target, msg.value, _gasLimit, _data);
    unchecked {
        ++nonce;
    }
}
\end{lstlisting}

To verify and finalize a withdrawal the following inputs are needed:
\begin{itemize}
    \item \verb|nonce|: the nonce of the message.
    \item \verb|sender|: the address of the sender on L2.
    \item \verb|target|: the address that is called on L1.
    \item \verb|data|: the data to be sent to the target.
    \item \verb|value|: the amount of ETH to be sent to the target.
    \item \verb|gasLimit|: the gas to be forwarded to the target.
    \item \verb|timestamp|: the timestamp of L2 corresponding to the output root.
    \item \verb|outputRootProof|: 4 bytes used to derive the output root.
    \item \verb|withdrawalProof|: proof of the withdraw inclusion.
    \item \verb|l2BlockNumber|: the number of the block where the withdraw was initiated.
\end{itemize}

The output root corresponding to the \verb|l2BlockNumber| is obtained from the  \\ \verb|L2OutputOracle|, it is verified that it is finalized, i.e., that the dispute period has passed, it is verified that the \verb|outputRootProof| matches the oracle root, it is verified that the hash of the withdrawal is included in it using the \verb|withdrawalProof|, that the withdrawal has not already been finalized, and the call to the \verb|target| is executed, sending \verb|gasLimit| amount, \verb|value| amount of Ether, and the \verb|data|.

\begin{lstlisting}[caption={Withdrawal finalization.},captionpos=b][language=Solidity]
function finalizeWithdrawalTransaction(
    Types.WithdrawalTransaction memory _tx,
    uint256 _l2BlockNumber,
    Types.OutputRootProof calldata _outputRootProof,
    bytes calldata _withdrawalProof
) external payable {
    require(
        l2Sender == DEFAULT_L2_SENDER,
        "OptimismPortal: can only trigger one withdrawal per transaction"
    );

    require(
        _tx.target != address(this),
        "OptimismPortal: you cannot send messages to the portal contract"
    );

    Types.OutputProposal memory proposal = L2_ORACLE.getL2Output(_l2BlockNumber);

    require(_isOutputFinalized(proposal), "OptimismPortal: proposal is not yet finalized");

    require(
        proposal.outputRoot == Hashing.hashOutputRootProof(_outputRootProof),
        "OptimismPortal: invalid output root proof"
    );

    bytes32 withdrawalHash = Hashing.hashWithdrawal(_tx);

    require(
        _verifyWithdrawalInclusion(
            withdrawalHash,
            _outputRootProof.withdrawerStorageRoot,
            _withdrawalProof\node{r}
    child { node {a} }
    child { node {b} };
        ),
        "OptimismPortal: invalid withdrawal inclusion proof"
    );

    require(
        finalizedWithdrawals[withdrawalHash] == false,
        "OptimismPortal: withdrawal has already been finalized"
    );

    finalizedWithdrawals[withdrawalHash] = true;

    require(
        gasleft() >= _tx.gasLimit + FINALIZE_GAS_BUFFER,
        "OptimismPortal: insufficient gas to finalize withdrawal"
    );

    l2Sender = _tx.sender;

    (bool success, ) = ExcessivelySafeCall.excessivelySafeCall(
        _tx.target,
        _tx.gasLimit,
        _tx.value,
        0,
        _tx.data
    );

    l2Sender = DEFAULT_L2_SENDER;

    emit WithdrawalFinalized(withdrawalHash, success);
}

\end{lstlisting}

\subsection{Cannon: the invalidity proof system}
\label{cannon}

If a Rollup validator, by locally executing batches and deposited transactions, discovers that the Layer 2 state does not match the root of the state published on-chain by the Sequencer, it can perform an invalidity proof on L1 to prove that the result of the block state transition is wrong.
Because of the overhead, processing an entire Rollup block on L1 is too expensive.
The solution is to execute on-chain only the first instruction of \verb|minigeth| of disagreement, compiling it into a MIPS architecture that is executed on an on-chain interpreter (of only 400 lines) and published on L1. \verb|minigeth| is a simplified version of geth in which the consensus, RPC, and database have been removed.

To find the disagreement instruction, an interactive binary search is conducted between the one who initiate the invalidity proof and the one who published the output root. When the proof starts, both parties publish the root of the MIPS memory state halfway through the execution of the block on the \verb|Challenge| contract: if the hash matches it means that both parties agree on the first half of the execution thus publishing the root of half of the second half, otherwise the half of the first half is published. Doing so achieves the first single instruction of disagreement in logarithmic time. If one of the two stops interacting, at the end of the dispute period the other participant automatically wins.

\subsubsection{Memory access}

The MIPS interpreter to process this instruction needs access to its memory: since the root is available, the necessary memory cells can be published by proving their inclusion. To access the state of the EVM, use is made of the \textit{Preimage Oracle}: given the hash of a block, it returns the block header, from which one can get the hash of the previous block and go back in the chain, or get the hash of the state and logs from which one can get the preimage. The oracle is implemented by \verb|minigeth| and replaces the database. Queries are made to other nodes to obtain the preimages.

\chapter{Validity Rollups}

The goal of a Validity Rollup is to formally demonstrate the validity of a state transition given the sequence of transactions to allow it to be verified in less time than would be spent processing them all entirely.

They differ according to their degree of compatibility with EVM \cite{vitalik2022zkevm}:
\begin{itemize}
    \item \textbf{Type 1}: the Rollup is totally equivalent to Ethereum and does not make any changes to facilitate proof generation. The advantage is that each tool is immediately reusable, but since Ethereum is not meant to be proven by validity proofs, proving a single block can take hours. Currently one research team is trying to implement a Type 1 Validity Rollup \cite{perez2022zkevm}.
    \item \textbf{Type 2}: the Rollup is totally equivalent to the EVM, but has differences from Ethereum such as some data structures, e.g., block structure or state tree. Ethereum clients are not reusable without modification but most applications would still work. The advantage is that the proof time decreases compared to Type 1, but most of the complexity is caused by the EVM itself. Scroll \cite{scroll2022} and Polygon Hermez \cite{hermez2022} are building this type of Rollup.
    \item \textbf{Type 3}: the Rollup is almost equivalent to the EVM with some differences to increase the speed of the prover. No one is currently building a Type 3 Rollup, but it is possible that Scroll and Hermez will be released initially in this manner and then achieve equivalence over time.
    \item \textbf{Type 4}: the Rollup uses a different virtual machine than the EVM, but it is possible to compile a language from the EVM into a language for this virtual machine. This provides maximum proof speed at the expense of compatibility with Ethereum tools. ZKSync \cite{zksync2022} and StarkNet \cite{starkware2022} are two examples.
\end{itemize}

\section{Preliminary}

In general, given a function, an input and an output the scope is to create a certificate that proves that the output is obtained by computing the function on the input and that verification of the certificate costs less than the cost of the function. These certificates are called \textit{computational integrity proofs} \cite{ben2017computational}. The property of a solution to be easily verifiable but difficult to compute is well known in the study of complexity, as is the case with NP-complete problems. With computational integrity proofs, an attempt is made to exploit this asymmetry even for arbitrary computations: in 1991 the PCP theorem \footnote{from Probabilistic Checkable Proof} was proved, which states that for any proof in NP, if the prover performs only a polynomial amount of extra work, a verifier can validate the proof in poly-logarithmic time \cite{babai1991non}. PCPs are generalized to \textit{interactive oracle proofs} for problems beyond the NP class \cite{arnon2022pcp}. 

\subsection{Homomorphic cryptography}
Homomorphic encryption allows encryption of a value to which arithmetic operations can still be applied.

A group $\mathbb{G}$ is a set of elements and a binary operation (denoted here as $\cdot$) that has the following four properties:
\begin{itemize}
    \item \textbf{Closure}: $\forall a,b \in \mathbb{G} : a \cdot b \in \mathbb{G}$.
    \item \textbf{Associativity}: $\forall a,b,c \in \mathbb{G} : a \cdot (b \cdot c) = (a \cdot b) \cdot c$.
    \item \textbf{Identity}: there exists an element denoted $1_\mathbb{G}$ such that $\forall a \in \mathbb{G} : 1_\mathbb{G} \cdot a = a \cdot 1_\mathbb{G} = a$.
    \item \textbf{Invertibility}: $\forall a \in \mathbb{G}, \exists b \in \mathbb{G} : a \cdot b = 1_\mathbb{G}$. This element is denoted by $a^{-1}$.
\end{itemize}

If the group operation is also commutative, the group is called \textit{abelian}.
A group is called \textit{cyclic} if there exists an element of the group $g$, called \textit{generator}, such that all the elements of the group can be written as $g^i$ for some positive integer $i$. \footnote{where $g^i$ denotes $\underbrace{g \cdot g \cdot \dots \cdot g}_\text{$i$ copie di $g$}$.} Every cyclic group is abelian. A subgroup $\mathbb{H}$ of $\mathbb{G}$ is a subset of $\mathbb{G}$ that forms a group under the same operation as $\mathbb{G}$. The cardinality $|\mathbb{G}|$ is called the \textit{order} of $\mathbb{G}$. By Lagrange's Theorem, the order of a subgroup $\mathbb{H}$ of $\mathbb{G}$ divides the order of $\mathbb{G}$.

Given a group $\mathbb{G}$, the discrete logarithm problem takes two elements of the group $a$ and $b$ and returns a positive integer $i$ such that $a^i = b$. If $\mathbb{G}$ is cyclic it is guaranteed that this $i$ exists. It is believed that finding the discrete logarithm is computationally intractable, but it has been shown that a quantum computer can solve the problem in polynomial time using Shor's algorithm \cite{shor1994algorithms}. The fastest known classical algorithm used in practice operates in time $O(\sqrt{|\mathbb{G}|})$ \cite{pollard1978monte}.

In cryptography, the groups used are typically cyclic subgroups of groups defined using elliptic curves over finite fields, or multiplicative groups of integers modulo a large prime number.

Given a generator $g$ of a cyclic group, it is then possible to encrypt a value $a$ by calculating $h = g^a$. Given this scheme, it is possible to multiply an encrypted value by a known value $b$ by calculating $h^b$, keeping the result encrypted:
$$h^b = (g^a)^b = g^{a \cdot b}$$
In addition, it is possible to sum two encrypted values using multiplication:
$$g^a \cdot g^b = g^{a+b}$$
and similarly subtract two encrypted values using division:
$$\frac{g^a}{g^b} = g^{a-b}$$

It is not possible to multiply or divide two encrypted values with each other, and it is not possible to calculate the power of an encrypted value.

\subsection{Probabilistic proofs}

Probabilistic proofs can be used to drastically improve the efficiency of some algorithms.
For example, Freivalds \cite{freivalds1977probabilistic} demonstrated how to verify that a matrix $C$ is the product of two matrices $A$ and $B$ of size $n \times n$ in time $O(n^2)$ (only one more constant than reading the matrices), while the fastest algorithm known to multiply two matrices is performed in about time $O(n^{2.37286})$ \cite{le2014powers} \cite{alman2021refined}. 

The technique used is based on converting the information into a vector of length $n$, interpreting the elements as coefficients of a polynomial, encode the vector as evaluations of the polynomial in a field $\mathbb{F}_p$ where $p \gg n$ for each $r \in \mathbb{F}_p$ and exploit the fact that two different polynomials of degree $n$ equal each other on at most $n-1$ points, allowing their equality to be probabilistically verified using sampling. This property can be extended to multi-variable polynomials to reduce the degree of the polynomial \cite{schwartz1980fast}. Since encodings are interpreted as evaluations, it is not necessary to construct the entire vector. This technique is called Reed-Solomon encoding \cite{wicker1999reed}, in which the distance amplification property is exploited. The encoding can alternatively be constructed by interpreting the elements of the original vector as the valuations and obtaining the polynomial using polynomial interpolation techniques such as Lagrange interpolation, with the advantage that the original vector is a sub-vector of the extended one.

\begin{theo}[Probabilistic comparison of two strings]{thm:stringscomparison}
    Alice and Bob want to determine whether the files they own are equal by minimizing the amount of information exchanged during communication. The files consist of a sequence of $n$ ASCII characters, so $m = 128$ possible characters. Alice owns the file $(a_1, \dots, a_n)$, while Bob owns the file $(b_1, \dots, b_n)$. The trivial solution is to send all $n$ characters, but this is not possible if $n$ is very large. There is no deterministic procedure that sends less information \cite{kushilevitz1997communication}, so they decide to use a probabilistic procedure. One sets a prime number $p \ge \max{\{m, n^2\}}$ for the field $\mathbb{F}_p$. For the rest of the example we assume that all operations are performed in this field. We define the family of hash functions $\mathcal{H} = \{h_r : r \in \mathbb{F}_p\}$ where $h_r(a_1, \dots, a_n) = \sum^n_{i=1} a_i \cdot r^{i-1}$, i.e., the result of evaluating the polynomial of degree $n-1$ over $r$ obtained by interpreting the characters as coefficients. Alice chooses a random element $r \in \mathbb{F}_p$, computes $v = h_r(a)$ and sends $r$ and $v$ to Bob. Bob checks whether $v = h_r(b)$, if so gives EQUAL in output otherwise NON-EQUAL.
    
    \subsubsection*{Completeness and soundness}
    Trivially, if $\forall i \in \{1, \dots, n\} : a_i = b_i$, then Bob gives in output EQUAL for every choice of $r$. If there is at least one $i$ for which $a_i \neq b_i$, then Bob outputs NON-EQUAL with probability at least $1-(n-1)/p$, that is, at least $1-1/n$ for $p \geq n^2$. To prove this, let $p_a(x) = \sum^n_{i=1} a_i \cdot x^{i-1}$ and similarly $p_b(x) = \sum^n_{i=1} b_i \cdot x^{i-1}$: if there is at least one $a_i \neq b_i$, then there are at most $n-1$ values of $r$ such that $p_a(r) = p_b(r)$. Since $r$ is randomly chosen by $\mathbb{F}_p$, the probability that Alice will draw such an $r$ is at most $(n-1)/p$, so the probability that he will output NON-EQUAL is at least $1-(n-1)/p$.
    
    \subsubsection{Protocol cost}
    The deterministic procedure has a cost of $n\log m$ bits exchanged. In the probabilistic procedure, only two elements of $\mathbb{F}_p$ are exchanged, namely $v$ and $r$: assuming $p \leq n^c$ for some constant $c$, the cost is $O(\log n)$.
\end{theo}

\subsection{Interactive proofs}

Interactive proofs were first formally introduced by Babai \cite{babai1985trading}.

Given a function $f:\{0,1\}^n \to R$ where $R$ is a finite interval, an interactive $k$-message \textit{proof system} for $f$ consists of a probabilistic algorithm $V$, called \textit{Verifier} that is executed in polynomial time and a deterministic algorithm $P$ called \textit{Prover}.
Both $V$ and $P$ are given a common input $x \in \{0,1\}^n$ and at the beginning of the protocol $P$ produces a value $y$ that it claims is equal to $f(x)$. Next, $P$ and $V$ exchange a series of $k$ messages alternately: at the end the Verifier will respond with $1$ or $0$ depending on whether it accepts the Prover's claim that $y = f(x)$ or not. The Verifier can be made deterministic by fixing the source of internal randomness a priori. We denote by $\text{out}(V,x,r,P) \in \{0,1\}$ the output of the Verifier $V$ on input $x$ interacting with $P$ with internal random value $r$.

An interactive proof system is said to possess completeness error $\delta_C$ and soundness error  $\delta_S$ if the following properties hold:
\begin{enumerate}
    \item \textbf{Completeness}: for each $x \in \{0,1\}^n$ and $P$ honest, 
    $$
    \mathbb{P}[\text{out}(V,x,r,P)=1] \ge 1 - \delta_C
    $$
    \item \textbf{Soundness}: for each $x \in \{0,1\}^n$ and for each $y \neq f(x)$ sent by $P'$ at the beginning of the protocol,
    $$
    \mathbb{P}[\text{out}(V,x,r,P')=1] \le \delta_S
    $$
\end{enumerate}

By convention, an interactive proof system is said to be valid if $\delta_C,\delta_S \le 1/3$.
In addition to the cost in time, determinant is the cost in space, the number of bits communicated, the number of messages sent, and whether $r$ is a value made public or not.

The interactive proofs illustrated below will have perfect completeness, i.e., $\delta_C = 0$, while the soundness error will be proportional to $1/|\mathbb{F}|$ where $\mathbb{F}$ is the field over which the proof is defined. At the practical level, a field will be chosen such that the soundness error is extremely small ($\le 2^{-128}$). This error can be further reduced to $\delta_S^k$ by repeating the protocol $k$ times.

Interactive proofs can also be adapted to languages by requiring that $V$ accept with high probability a word from the language and reject with high probability a word not contained in the language. The main difference to using languages instead of functions is that a non-inclusion proof is not required for non-included words. 

Given a function $f$, an interactive proof for $f$ is equivalent to an interactive proof for the language $\mathcal{L}_f \coloneqq \{(x,y): y = f(x)\}$.

\subsubsection*{IP = PSPACE}
Let IP be the class of languages whose membership is provable using an interactive proof system with a Verifier operating in polynomial time. The class NP is a subset of IP in that it can be seen as a restriction of it in which the demonstrations are deterministic and non-interactive, that is, having zero completeness and soundness error. Adi Shamir \cite{shamir1992ip} succeeded in characterizing the IP class by proving its equivalence to PSPACE, which is believed to be much larger than NP.

\subsection{Zero-Knowledge proofs}

Zero-knowledge proofs are a type of probabilistic proof between a Prover and a Verifier first introduced by Goldwasser et al. \cite{goldwasser1989knowledge}.
Given a proposition $x$, there may exist a \textit{witness} $w$ that proves a relation $R$ between $x$ and $w$ to a Verifier $V$. For example, $x$ may be a hash, and $w$ the preimage such that the relation $x = H(w)$ holds, where $H$ is a hash function. It is said that $x$ is true if and only if there exists $w$ such that $R(x, w)$ holds. The Prover $P$ wants to prove to $V$ that $x$ is true.

The three properties of a zero-knowledge proof are:
\begin{itemize}
    \item \textbf{Completeness}: if $R(x, w)$ is true, then $P(x, w) \iff V(x)$: if $P$ possesses a witness that proves the proposition, then he will be able to convince $V$ that the proposition is true. Important to note that the Verifier does not take the witness as input.
    \item \textbf{Soundness}: if $x$ is false, $V(x)$ will reject (with high probability) for every $P$.
    \begin{itemize}
        \item \textbf{Knowledge soundness}: if $P$ makes $V(x)$ accept, then there is a way to extract $w$ such that $R(x, w)$. This stronger form of soundness is often required because there are propositions that are trivially always true, such as the existence of a preimage of a 256-bit hash for SHA256.
    \end{itemize}
    \item \textbf{Zero-Knowledge}: if $R(x, w)$ holds, then we can simulate the execution of the proof using only $x$. This shows that the Verifier is unable to extract any information about $w$.
\end{itemize}

\begin{theo}[Schnorr Protocol]{thm:schnorr}

The Schnorr \cite{schnorr1989efficient} protocol is a very simple use of zero-knowledge proofs in that they are applied to a specific problem and the technique is not general.
Let $p$ be a prime number and let $g$ be a publicly known generator of a cyclic group $\mathbb{G}$ of prime order $q$. To generate a key pair, Alice:
\begin{itemize}
    \item Chooses a random number $a$ from $1$ to $q$.
    \item Calculates the public key $\mbox{PK}_A = g^a \mod p$.
    \item Saves the secret key $\mbox{SK}_A = a$.
\end{itemize}

Alice wants to prove to Bob that she knows the secret key (the witness $w$) that corresponds to the public key using an interactive protocol:

\begin{enumerate}
    \item Alice chooses a random $k$ from $1$ to $q$ and calculates $h = g^k \mod p$ and sends it to Bob.
    \item Bob chooses a random $c$ and sends it to Alice.
    \item Alice responds with $s = ac+k \mod q$.
    \item Bob checks that $g^s \equiv \mbox{PK}_A^c \cdot h \mod p$. 
\end{enumerate}

The protocol is a zero-knowledge proof in that it meets the three properties listed earlier:
\begin{itemize}
    \item \textbf{Completeness}:
    \begin{align*}
        g^s &\equiv PK_A^c \cdot h \mod p\\
        g^{ac+k} &\equiv (g^a)^c \cdot g^k \mod p\\
        g^{ac+k} &\equiv g^{ac+k} \ \ \ \ \ \mod p
    \end{align*}
    \item \textbf{Knowledge soundness}: to prove that Alice actually knows the secret key $a$, we use a special type of Verifier called \textit{knowledge extractor}, which can extract the key from the proof. This does not contradict the zero-knowledge property since the extractor is not allowed to exist if the proof is done correctly. The extractor, after the third step, restarts the protocol from step 2, making Alice use the same blinding factor $k$. With this strategy it is possible to extract $\text{SK}_A$:
    \begin{align*}
        & \ \ \ \ (s_1 - s_2) / (c_1 - c_2) \ \ \ \ \ \ \ \ \ \ \ \ \ \mod q \\
        &= (ac_1+k-ac_2-k)/(c_1-c_2) \mod q\\
        &= a(c_1-c_2)/(c_1-c_2) \ \ \ \ \ \ \ \ \ \ \ \ \mod q\\
        &= a
    \end{align*}
    \item \textbf{Zero-Knowledge}: the goal is to prove that one knows the secret $a$ for some public key $g^a \mod p$ without knowing $a$, using a special type of Prover called \textit{Simulator}. Alice sends Bob $g^{k_1}$ to find out which random $c$ Bob chooses. After that she rewinds the protocol, this time choosing $g^{k_2} = g^z * g^{a(-c)}$ as the initial message. When Bob resends the $c$ challenge, Alice responds with $z$. Since from the Verifier's point of view this protocol is identical to the real protocol, she does not reveal any information about $a$. This does not conflict with the correctness property since the Simulator is not allowed to exist if the proof is carried out correctly.
\end{itemize}
\end{theo}

\subsection{SNARK}

A SNARK (Succint Non-interactive ARgument of Knowledge) proof is a category of probabilistic proof by arbitrary computation. The main characteristics are constant proof size and constant verification time. Their non-interactivity is given by a trusted setup to generate the source of randomness.

To construct a SNARK, the steps to be performed are:
\begin{enumerate}
    \item Conversion of the computation into an arithmetic circuit.
    \item Conversion of the circuit into a Rank-1 Constraint System (R1CS).
    \item Construction of the Quadratic Arithmetic Program (QAP).
    \item SNARK proof construction.
\end{enumerate}

For the purpose of illustration, suppose we need to demonstrate the correct execution of the following program for some $x$:
\begin{minted}
[
frame=lines,
framesep=2mm,
baselinestretch=1.2,
fontsize=\footnotesize,
linenos
]
{python}
def f(x):
    y = x**3
    return y + 8
\end{minted}

which calculates the result of the polynomial $x^3+8$.

\subsubsection*{Arithmetic circuits}
An arithmetic circuit takes numeric signals as input and applies multiplications and additions to them in a finite prime field. An arithmetic circuit can have multiple intermediate signals and an output signal. One of the most widely used languages for defining arithmetic circuits is Circom.

The code under consideration must be converted into one that uses only assignment and operations of the form $x = y @ z$ where $@ \in (+,-,*,/)$ in a process called \textit{flattening}.
The corresponding flattened code is thus:
\begin{minted}
[
frame=lines,
framesep=2mm,
baselinestretch=1.2,
fontsize=\footnotesize,
linenos
]
{python}
def f(x):
    n = x*x
    m = n*x
    out = m + 8
\end{minted}

\begin{figure}
    \centering
    \begin{forest}
for tree={l sep=30pt, s sep=50pt}
[,
    [+, fill=lime, edge={<-}, edge label={node[midway,fill=white] {out}}
      [\textasteriskcentered{}, fill=lime, edge={<-}, edge label={node[midway,fill=white] {m}}
        [\textasteriskcentered{}, fill=lime, edge={<-}, edge label={node[midway,fill=white] {n}}
            [x, tier=word, fill=yellow, edge={<-}]
            [x, tier=word, fill=yellow, edge={<-}]
        ]
        [x, tier=word, fill=yellow, edge={<-}]
      ]
        [8, edge={<-}, tier=word, fill=pink] 
    ]
]
\end{forest}
    \caption{The program represented as an arithmetic circuit.}
    \label{fig:my_label}
\end{figure}
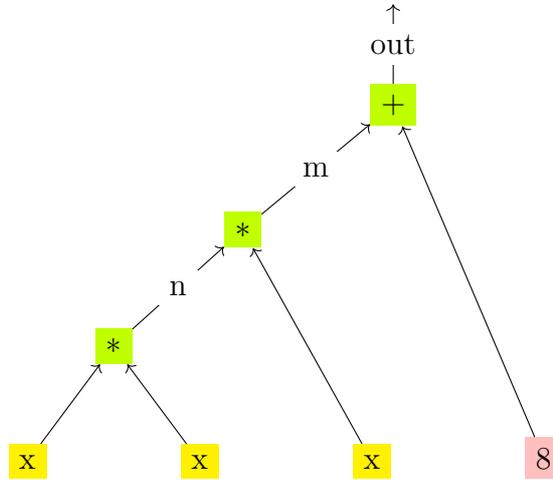

\subsubsection*{R1CS}
The gates of an arithmetic circuit are represented as a set of equations of a Rank-1 Constraint System (R1CS). An R1CS is a sequence of sets of three vectors $(a_i,b_i,c_i)$ called constraints and a solution is a vector $s$ such that $\forall i : s \cdot a_i * s \cdot b_i - s \cdot c_i = 0$. The length of the vectors is given by the number of variables in the system plus the number $1$ to express the constants and a variable \verb|out| represented the output. Each gate of the circuit is represented by a constraint.

The vector $s$ corresponding to the program under consideration is:
$$
s = \begin{bmatrix}
    1 \\ 
    x \\
    n \\
    m \\
    \text{out}
    
\end{bmatrix}
$$

A constraint is then defined for each line of the flattened program. 
The following group is defined for line 2:
$
a_1 = [0 \ 1 \ 0 \ 0 \ 0], \
b_1 = [0 \ 1 \ 0 \ 0 \ 0] , \
c_1 = [0 \ 0 \ 1 \ 0 \ 0] \
$.
Note that the constraint $s \cdot a_1 * s \cdot b_1 - s \cdot c_1 = 0$ is met.

The other constraints are:
\begin{itemize}
    \item Line 3: $
a_2 = [0 \ 0 \ 1 \ 0 \ 0], \
b_2 = [0 \ 1 \ 0 \ 0 \ 0] , \
c_2 = [0 \ 0 \ 0 \ 1 \ 0] \
$.
    \item Line 4: $
a_3 = [8 \ 0 \ 0 \ 1 \ 0], \
b_3 = [1 \ 0 \ 0 \ 0 \ 0] , \
c_3 = [0 \ 0 \ 0 \ 0 \ 1] \
$.
\end{itemize}

Running the program is equivalent to finding a $s$ such that it satisfies the R1CS. An example is $s = [1 \ 3 \ 9 \ 27 \ 35]$.

\subsubsection*{QAP}

The next step is to convert the vectors into polynomials $A_i(x)$, $B_i(x)$ and $C_i(x)$ for $i \in [1,N]$ where $N$ is the number of elements in the constraints. These polynomials are constructed by requiring that $A_i(n) = a_{n,i}$, $B_i(n) = b_{n,i}$ and $C_i(n) = c_{n,i}$ and using Lagrange interpolation. In our case we go from 3 groups of 3 vectors of length 5 to 5 groups of 3 polynomials of degree 2.

For example, for $A_1$ we have $A_1(1) = 0$, $A_1(2) = 0$, $A_1(3) = 8$ and interpolating we get $A_1(x) = 4x^2 -12x + 8$.

Similarly for the other polynomials:
\begin{itemize}
    \item $A_2(x) = \frac{1}{2}x^2-\frac{5}{2}x+3$
    \item $A_3(x) = -x^2+4x-3$
    \item $A_4(x) = \frac{1}{2}x^2-\frac{3}{2}x+1$
    \item $A_5(x) = 0$
    \item $B_1(x) = \frac{1}{2}x^2-\frac{3}{2}x+1$
    \item $B_2(x) = -\frac{1}{2}x^2+\frac{3}{2}x$
    \item $B_3(x) = 0$
    \item $B_4(x) = 0$
    \item $B_5(x) = 0$
    \item $C_1(x) = 0$
    \item $C_2(x) = 0$
    \item $C_3(x) = \frac{1}{2}x^2-\frac{5}{2}x+3$
    \item $C_4(x) = -x^2+4x-3$
    \item $C_5(x) = \frac{1}{2}x^2-\frac{3}{2}x+1$
\end{itemize}

Calculating for $x = 1$ gives the first set of constraints, for $x = 2$ the second set, and so on. For example, calculating for $x = 3$ gives:
$$
[A_1(3) \ A_2(3) \ A_3(3) \ A_4(3) \ A_5(3)] = [8 \ 0 \ 0 \ 1 \ 0] = a_3
$$
$$
[B_1(3) \ B_2(3) \ B_3(3) \ B_4(3) \ B_5(3)] = [1 \ 0 \ 0 \ 0 \ 0] = b_3
$$
$$
[C_1(3) \ C_2(3) \ C_3(3) \ C_4(3) \ C_5(3)] = [0 \ 0 \ 0 \ 0 \ 1] = c_3
$$

At this point the R1CS can be expressed as one equation:
$$
A(x) * B(x) - C(x) = H(x) * Z(x)
$$

where $A(x)$ is the scalar product between $s$ and $[A_1(x) \ A_2(x) \ A_3(x) \ A_4(x) \ A_5(x)]$ and similarly for $B(x)$ and $C(x)$. Since $\forall x \in [1,3] : P(x) = A(x)*B(x)-C(x)=0 $, the polynomial is a multiple of $Z(x) = \prod_{i=1}^3 (x-i)$. $H(x)$ is computed as $P(x)/Z(x)$.

In the presented example:
\begin{align*}
    P(x) &= -10x^4+54x^3-74x^2-6x+36 \\
    Z(x) &= x^3-6x^2+11x-6 \\
    H(x) &= -10x-6
\end{align*}

If $s$ is not a solution of R1CS then $P(x)$ is not divisible by $Z(x)$.

Using this property, the following naive protocol is developed:
\begin{enumerate}
    \item The Verifier generates a random value $r$, and knowing the number of constraints calculates $Z(r)$ and sends $r$ to the Prover.
    \item The Prover computes the polynomial $H(x)$, evaluates $P(r)$ and $H(r)$ and sends them to the Verifier.
    \item The Verifier checks that $P(r) = H(r)*Z(r)$. Since two different polynomials of degree $n$ equal each other on at most $n-1$ points, if the equality holds there is a high probability that the Verifier actually knows $P(x)$.
\end{enumerate}

One problem is that the Prover could use a random $h$ and compute a $p$ such that $p = h * Z(r)$ without actually knowing $P(x)$. Furthermore, no check is made on the degree of the polynomial.

A partial solution is to hide the value of $r$ using homomorphic encryption:
\begin{enumerate}
    \item The Verifier generates a random value $r$, evaluates $Z(r)$ and computes the encrypted values of the powers of $r$ needed for the polynomial, in our case $g^{r^0}$, $g^{r^1}$, $g^{r^2}$, $g^{r^3}$ and $g^{r^4}$ and sends them to the Prover. \footnote{because, as explained above, it is not possible to calculate the power of a cipher value}.
    \item The Prover computes $H(x)$, using the cipher values computes $g^{P(r)}$ and similarly $g^{H(r)}$ and sends them to the Verifier.
    \item The Verifier checks that $g^{P(r)} = \left(g^{H(r)}\right)^{Z(r)} = g^{H(r) \cdot Z(r)}$.
\end{enumerate}

In this way the Prover is restricted to using the selection of powers of $r$ given to him, but this restriction is not imposed: for example, given a random value $s$, the Prover can compute $z_h = g^s$ and $z_p = \left(g^{Z(r)}\right)^s$. The Verifier verifying
$$\left(g^{Z(r)}\right)^s = \left(g^s\right)^{Z(r)}$$
would erroneously accept the proof.

We obtain the desired result using the Knowledge-of-Exponent Assumption (KEA1) \cite{bellare2004knowledge} first introduced by Ivan Damgard \cite{damgaard1991towards}: informally, let $g$ be a generator, given $g$ and $g^a$, the only way to compute a pair $(C,Y)$ where $Y = C^a$ is to compute, choosing some $c$, $C = g^c$ and $Y = \left(g^a\right)^c$.

This assumption can be used to ask the Prover to compute results on $g^r$ and on a value \textit{shifted} $g^{\alpha r}$, used as ``checksum."
\begin{enumerate}
    \item The Verifier sends to the Prover the values $g^{r^0}$, $g^{r^1}$, $\dots$, $g^{r^d}$ and their shifts $g^{\alpha r^0}$, $g^{\alpha r^1}$, $\dots$, $g^{\alpha r^d}$.
    \item The Prover using these values calculates $g^p = g^{P(r)}$ and $g^{p'} = g^{\alpha P(r)}$.
    \item The Verifier verifies that $\left(g^p\right)^\alpha = g^{p'}$.
\end{enumerate}

This approach was first introduced by Jens Groth \cite{groth2010short}, creator of the Groth16 proof system \cite{groth2016size}.

\subsubsection*{Zero-Knowledge proof}

The zero-knowledge property can be attacked using a brute-force attack on the coefficients of the polynomial: the protocol must be secure even if there is only one coefficient and it is 1.

The checks that the Verifier performs are:
\begin{align*}
    g^{p} &= \left(g^{H(r)}\right)^{Z(r)} &\text{(verification of polynomial roots)} \\
    \left(g^p\right)^\alpha &= g^{p'} &\text{(verification of correct use of polynomial)}
\end{align*}

The zero-knowledge property can be preserved for brute force attacks by reapplying a shift on the values of a $\delta$ value. To extract information, the Verifier would have to find this value, which is considered computationally infeasible. Moreover, randomization is statistically indistinguishable from chance.

The Verifier's operations thus become:

\begin{align*}
    \left(g^{p}\right)^\delta &= \left(\left(g^{H(r)}\right)^\delta\right)^{Z(r)} \\
    \left(\left(g^p\right)^\delta\right)^\alpha &= \left(g^{p'}\right)^\delta
\end{align*}

\subsubsection*{Non-interactivity and trusted setup}

The only one who can be certain of the validity of an interactive zero-knowledge proof is the Verifier. From an outside observer's point of view, the Verifier may have colluded with the Prover by telling him the secret values $r$ and $\alpha$. This is useful in some applications where you do not want to allow the proof to be replicated to other \cite{jakobsson1996designated}, but in the case of distributed systems such as the blockchain it is inefficient to recreate the proof for everyone.

The parameters you want to keep secret are $Z(r)$ and $\alpha$. One could use the same method used to encrypt the powers of $r$, but as already mentioned homomorphic encryption does not allow multiplication of two cipher values.

The solution is the use of bilinear cryptographic \cite{dutta2004pairing} maps, i.e., a function $e(g^*, g^*)$ that given two cipher inputs $g^a$ and $g^b$ produces deterministically, using a map, their multiplied representation $e(g^a, g^b) = e(g,g)^{ab}$. Since the function uses two different groups as domain and co-domain, it is not possible to multiply the result by another cipher value. The main properties, obtained using elliptic curves, can be expressed as the following equations:
$$
    e(g^a,g^b) = e(g^b,g^a) = e(g^{ab}, g^1) = e(g^1, g^{ab}) = e(g^1, g^a)^b = e(g^1, g^1)^{ab}
$$

We assume there is a trusted participant who generates the secrets $r$ and $\alpha$ and after computing the cipher powers and their $\alpha$-shifts decides to eliminate the plaintext values. These parameters are called ``common reference string'' (CRS).
They are divided into two groups: 
\begin{itemize}
    \item \textbf{Proving key}: $\forall i \in \{0, \dots, d\}: (g^{r^i}, g^{\alpha r^i})$
    \item \textbf{Verification key}: $(g^{Z(r)}, g^\alpha)$
\end{itemize}

Using the verification key and having obtained $g^p, g^{p'}$ and $g^{H(r)}$ from the Prover, the Verifier checks:
\begin{itemize}
    \item $e(g^p, g^1) = e(g^{Z(r)}, g^{H(r)})$
    \item $e(g^p, g^\alpha) = e(g^{p'}, g^1)$
\end{itemize}

The problem with this approach is that one must trust that secret values, which are named ``toxic waste'', will be eliminated. Eli Ben-Sasson et al. described in 2015 a method to minimize this assumption by using a multiparty computation (MPC) \cite{ben2015secure}, in which an arbitrary number of participants contribute and at least one of them is required to delete their generated values to make the procedure secure. ZCash, an anonymous blockchain based on SNARK \cite{hopwood2016zcash}, organized the ``ceremony of powers of tau'' to generate its own common reference string.

\subsection{STARK}
\label{stark}

The STARK proof system was introduced in 2018 as a special case of SNARK \cite{ben2018scalable}. It uses hash functions as the only cryptographic assumption and the zero-knowledge property is optional. The time complexity of the Prover is $O(T \cdot \text{poly} \log(T))$ with respect to the time limit $T$ of the original computation, whereas in SNARK it depends on the implementation. The complexity in time for the Verifier is $O(\text{poly} \log(T))$. The two major advantages of STARK as an alternative to SNARK are the absence of a trusted setup and the resistance to quantum attacks, since no use is made of the discrete logarithm problem and elliptic curves to secure the system.

The proofs are made non-interactive by using the Fiat-Shamir transform \cite{fiat1986prove}: a hash function is used to generate the Verifier's random values, which takes as input the transcription of the protocol up to the point where these values are needed, so that it prevents it from choosing inputs that compute a hash suitable for generating valid but incorrect proofs.

\subsection*{Insights}

Martin Furer et al. \cite{furer1989completeness} showed that it is possible to convert any interactive demonstration system with completeness error $\delta_C \le 1/3$ to one with perfect completeness with Verifier polynomial blowup.

Goldwasser and Sipser \cite{goldwasser1986private} proved that it is possible to convert any interactive demonstration system with private randomness to one with public randomness with polynomial blowup for the Verifier.

\section{StarkNet}

\subsection{Overview}

StarkNet is a Validity Rollup developed by StarkWare that uses the STARK proof system to validate its state on Ethereum. To facilitate the construction of validity proofs, a virtual machine different than the EVM is used, whose high-level language is Cairo. For example, instead of \verb|keccak256| hashes, Pedersen hashes are used where possible.

\subsection{Deposits}
Users can deposit transactions via a contract on Ethereum by calling the \verb|sendMessageToL2| function.

\begin{lstlisting}[language=Solidity]
function sendMessageToL2(
    uint256 toAddress,
    uint256 selector,
    uint256[] calldata payload
) external override returns (bytes32) {
    uint256 nonce = l1ToL2MessageNonce();
    NamedStorage.setUintValue(L1L2_MESSAGE_NONCE_TAG, nonce + 1);
    emit LogMessageToL2(msg.sender, toAddress, selector, payload, nonce);
    bytes32 msgHash = getL1ToL2MsgHash(toAddress, selector, payload, nonce);
    l1ToL2Messages()[msgHash] += 1;

    return msgHash;
}
\end{lstlisting}

The message is recorded in the contract by computing its hash and increasing a counter. Sequencers listen for the \verb|LogMessageToL2| event and encode the information in a StarkNet transaction that calls a function of a contract that has the decorator \verb|l1_handler|. At the end of execution, when the proof of state transition is produced, the consumption of the message is attached to it and it is deleted by decreasing its counter.

Below is an example of \verb|l1_handler| for depositing fungible tokens:
\begin{lstlisting}[language=Cairo]
@l1_handler
func deposit{
    syscall_ptr : felt*,
    pedersen_ptr : HashBuiltin*,
    range_check_ptr,
}(from_address : felt, user : felt, amount : felt):
    # Make sure the message was sent by the intended L1 contract.
    assert from_address = L1_CONTRACT_ADDRESS

    # Read the current balance.
    let (res) = balance.read(user=user)

    # Compute and update the new balance.
    tempvar new_balance = res + amount
    balance.write(user, new_balance)

    return ()
end
\end{lstlisting}

A contract can have multiple \verb|l1_handlers| that are differentiated by a selector, which can be obtained using the StarkNet Python library:

\begin{lstlisting}[language=Python]
from starkware.starknet.compiler.compile import get_selector_from_name

print(get_selector_from_name('deposit'))
\end{lstlisting}

\subsubsection*{Gas market}

The inclusion of deposited transactions is not required by the StarkNet specification, so a gas market is needed to incentivize Sequencers to publish them on L2. In the current version, because the Sequencer is centralized and managed by StarkWare, the cost of deposited transactions is only determined by the cost of executing the deposit. Since StarkNet version 0.10.0, the cost of execution on L2 is paid by sending ETH to \verb|sendMessageToL2| which becomes \verb|payable|, and the \verb|LogMessageToL2| event adds a \verb|uint256 fee| field. These ETHs remain locked on L1 and are transferred to the Sequencer, also on L1, when the deposited transaction is included in a state transition. The amount of ETH sent, if the deposited transaction is included, is fully spent, regardless of the amount of gas consumed on L2.

\subsubsection*{L1 Attributes Deposited Transaction}
StarkNet does not have a system that makes L1 block attributes available automatically. Alternatively, Fossil is a protocol developed by Oiler Network \cite{oiler2022fossil} that allows, given a hash of a block, any information to be obtained from Ethereum by publishing preimages.

\subsection{Sequencing}
\label{starknetDA}

The current state of StarkNet can be derived entirely from Ethereum. Any state difference between transitions is published on L1 as calldata.

\subsubsection*{Sequencer}
Differences are published for each contract and are saved as \verb|uint256[]| and follow the following coding:
\begin{itemize}
    \item Number of fields concerning contract publications.
    \item For each published contract:
    \begin{itemize}
        \item \verb|contract_address|: the address of the published contract.
        \item \verb|contract_hash|: the hash of the published contract.
        \item \verb|len(constructor_call_data)|: the number of arguments of the contract constructor.
        \item \verb|constructor_call_data|: list of constructor arguments.
    \end{itemize}
    \item Number of contracts whose storage has been modified.
    \item For each contract that has been modified:
    \begin{itemize}
        \item \verb|contract_address|: the address of the modified contract.
        \item \verb|num_of_storage_updates|: number of storage modifications.
        \item \verb|key, value|: pair of the storage address of the contract whose value was changed and the new value.
    \end{itemize}
\end{itemize}

\subsubsection*{Derivation}
The state differences are published in order, so it is sufficient to read them sequentially to reconstruct the state. The derivation is implemented by Pathfinder, a full node client of StarkNet.

\subsection{Withdrawals}

To send a message from L2 to L1, the syscall \verb|send_message_to_L1| is used. The message is published to L1 by increasing its hash counter along with the proof and finalized by calling the function \verb|consumeMessageFromL2| on L1, which decrements the counter.

\begin{lstlisting}[language=Solidity]
function consumeMessageFromL2(uint256 fromAddress, uint256[] calldata payload)
    external
    override
    returns (bytes32)
{
    bytes32 msgHash = keccak256(
        abi.encodePacked(fromAddress, uint256(msg.sender), payload.length, payload)
    );

    require(l2ToL1Messages()[msgHash] > 0, "INVALID_MESSAGE_TO_CONSUME");
    emit ConsumedMessageToL1(fromAddress, msg.sender, payload);
    l2ToL1Messages()[msgHash] -= 1;
    return msgHash;
}
\end{lstlisting}

The official StarkNet bridge is the StarkGate, and it uses the messaging system in the following way to transfer ETH from L2 to L1:

\begin{lstlisting}[language=Solidity]
function withdraw(uint256 amount, address recipient) public override {
    consumeMessage(amount, recipient);
    // Make sure we don't accidentally burn funds.
    require(recipient != address(0x0), "INVALID_RECIPIENT");
    require(address(this).balance - amount <= address(this).balance, "UNDERFLOW");
    recipient.performEthTransfer(amount);
}
\end{lstlisting}

\begin{lstlisting}[language=Solidity]
function consumeMessage(uint256 amount, address recipient) internal {
    emit LogWithdrawal(recipient, amount);

    uint256[] memory payload = new uint256[](4);
    payload[0] = TRANSFER_FROM_STARKNET;
    payload[1] = uint256(recipient);
    payload[2] = amount & (UINT256_PART_SIZE - 1);
    payload[3] = amount >> UINT256_PART_SIZE_BITS;

    messagingContract().consumeMessageFromL2(l2TokenBridge(), payload);
}
\end{lstlisting}

Since in the call of \verb|consumeMessageFromL2| the caller is the StarkGate contract, anyone can finalize any withdrawal.

\subsection{Validity proofs}

The Cairo Virtual Machine \cite{goldberg2021cairo} is designed to facilitate the construction of STARK proofs. The Cairo language allows the computation to be described as a program and not directly as a circuit \footnote{as is the case, for example, with the Circom language}. This is accomplished by using a system of polynomial equations (AIR) representing a single computation: the FDE cycle of a von Neumann architecture. The number of constraints is thus fixed and independent of the type of computation, allowing for only one Verifier for every program.

The instruction set is defined as an \textit{Algebraic RISC}, in which addition and multiplication operations are performed over a finite prime field; verification of the equality of two values is supported but comparison if one value is less than the other is not. This trade-off arises from an evaluation around the fact that adding an extra instruction to the instruction set increases the complexity of executing a single step but can reduce the number of steps in a program.

The architecture possesses three registers:
\begin{itemize}
    \item \textbf{ap}: \verb|allocation pointer|, points to the first unused memory cell.
    \item \textbf{fp}: \verb|frame pointer|, points to the current function frame. The memory address of the function arguments and local variables are relative to this value.
    \item \textbf{pc}: \verb|program counter|, points to the current instruction.
\end{itemize}

To increase the efficiency of programs, use is made of \textit{nondeterministic programming}. Suppose we want to compute the square root of a certain number $x$: the deterministic approach is to use an algorithm that computes $y = \sqrt{x}$ and consequently include the computation in the proof; in the nondeterministic approach, the Prover computes $y = \sqrt{x}$ with the same algorithm but without including it in the proof, but including the verification of $y^2 = x$, which can be carried out with a single instruction. From the Verifier's point of view, the value of $y$ is ``guessed", and the only computation that is performed is its verification. In this particular case, $-y$ is also a value accepted by the Verifier, despite the fact that the deterministic version returns only $y$. These nondeterministic instructions are called \textit{hints}, and since they are executed only by the Prover they can be written in any language. The current implementation of the Cairo language uses Python to write hints.

The memory is read-only and nondeterministic: the Prover chooses all memory values, which cannot be changed. Another requirement, for the sake of efficiency, is that it be contiguous. To calculate the square root of $25$, in Cairo one can write:
\begin{lstlisting}[language=Cairo]
[ap] = 25; ap++
[ap - 1] = [ap] * [ap]; ap++
\end{lstlisting}
that is: the cell pointed to by \verb|ap| must contain $25$, and the next cell must contain a value $x$ that satisfies $25 = x*x$. In order to generate the proof, the Prover must assign values to memory in such a way that these constraints are satisfied. This can be done using hints:
\begin{lstlisting}[language=Cairo]
[ap] = 25; ap++
%{
    import math
    memory[ap] = int(math.sqrt(memory[ap - 1]))
%}
[ap - 1] = [ap] * [ap]; ap++
\end{lstlisting}

The input of a program is its witness, while the output is the data to be shared with the Verifier. In the current version it is represented as a JSON file. Assuming you have a program that given $n$ calculates the $n$-th Fibonacci number $y$, if the output includes both $n$ and $y$, then you are proving to the Verifier that $y$ is the $n$-th Fibonacci number, if the output is only $y$, you are proving to the Verifier that you know an $n$ such that the $n$-th Fibonacci number is $y$, whereas if the output is only $n$ you are proving that you have computed the $n$-th Fibonacci number without showing the result.

Formally we define two Cairo machines, one deterministic, used by the Prover, and one nondeterministic used by the Verifier. We fix a prime field $\mathbb{F}_P = \mathbb{Z}/P$ and a finite extension $\mathbb{F}$ of it:
\newtheorem{definition}{Definition}
\begin{definition}
The Cairo machine is a function that receives the following inputs:
\begin{itemize}
    \item a number of steps $T \in \mathbb{N}$
    \item the memory function $m:\mathbb{F} \to \mathbb{F}$
    \item a sequence $S$ of $T+1$ states $S_i=(\text{pc}_i,\text{ap}_i,\text{fp}_i) \in \mathbb{F}^3$ for $i \in [0,T]$
\end{itemize}
and the output is "accept" or "reject". The machine accepts if and only if for each $i$ the state transition from state $i$ to state $i+1$ is valid.
\end{definition}

The decision whether a single transition is valid or not depends only on the two states $S_i$ and $S_{i+1}$ under consideration. The memory function $m$, since in practice $|\mathbb{F}|$ is very large, can be seen as a sparse function in which almost all values are zero.

\begin{definition}
The nondeterministic Cairo machine receives the following inputs:
\begin{itemize}
    \item a number of steps $T \in \mathbb{N}$
    \item the partial memory function $m^*:A^* \to \mathbb{F}$, where $A^* \subseteq \mathbb{F}_P$
    \item initial and final values of the registers \verb|pc|, \verb|ap|, and \verb|fp|.
\end{itemize}

and the output is "accept" or "reject." The machine accepts if and only if there exists a memory function $m:\mathbb{F} \to \mathbb{F}$ that extends $m^*$ and a list of states $S_i \in \mathbb{F}^3$ for $i \in [0,T]$ such that the initial and final states correspond to the input values and that a deterministic Cairo machine accepts.
\end{definition}

Computing whether or not the deterministic version accepts an input can be computed in polynomial time by a deterministic machine, while the nondeterministic version needs a nondeterministic machine to be computed in polynomial time.

The bytecode of a program is a sequence of field elements $b = (b_0, \dots, b_{|b|-1})$ and two indices $\text{prog}_\text{start}$, $\text{prog}_\text{end} \in [0, |b|]$. To run the program, we choose an index $\text{prog}_\text{base} \in \mathbb{F}$, you set the partial memory function $m^*$ such that $\forall i \in [0, |b|]: m^*(\text{prog}_\text{base}+i)=b_i$ and assigns the initial value of \verb|pc| to $\text{prog}_\text{base}+\text{start}$ and the final value to $\text{prog}_\text{base}+\text{end}$. In addition, the partial function might contain other assignments that add additional constraints, such as input arguments.

The Cairo runner is responsible for executing these programs. The main difference between executing a normal program and a program in Cairo is that the latter, as seen above, allows nondeterministic instructions. The Cairo runner uses hints to infer these values.
The output of the runner consists of:
\begin{itemize}
    \item an input that the nondeterministic Cairo machine accepts:
    $$(T, m^*, \verb|pc|_I, \verb|pc|_F, \verb|ap|_I, \verb|ap|_F)$$
    where $m^*$ contains the bytecode of the program, additional information such as hints, $\verb|pc|_I = \text{prog}_\text{base}+\text{prog}_\text{start}$ and $\verb|pc|_F = \text{prog}_\text{base}+\text{prog}_\text{end}$.
    \item an input that the deterministic Cairo machine accepts:
    $$(T, m, S)$$
\end{itemize}
The runner fails if the execution results in a contradiction or if it fails to compute some values due to insufficient hints. The STARK prover uses a deterministic input to generate a proof that  the nondeterministic input is accepted by the nondeterministic Cairo machine. 

StarkNet aggregates multiple transactions into a single STARK proof using a shared prover named SHARP. The proofs are sent to a smart contract on Ethereum, which verifies their validity and updates the Merkle root corresponding to the new state. The sub-linear cost of verifying a proof of validity allows its cost to be amortized over multiple transactions.

\chapter{Comparison}

\section{Withdrawal time}

The most important aspect that distinguishes Optimistic Rollups from Validity Rollups is the time that elapses between the initialization of a withdrawal and its finalization.

In both cases, withdrawals are initialized on L2 and finalized on L1. On StarkNet, finalization is possible as soon as the validity proof of the state root is accepted on Ethereum: theoretically, it is possible to withdraw funds in the first block of L1 following initialization, following verification of validity. In practice, the frequency of sending validity proofs on Ethereum is a trade-off between the speed of block finalization and proof aggregation. Currently StarkNet provides validity proofs for verification every 10 hours \cite{starknet2022frequency}, but this interval will be decreased as transactions increase.

On Optimism Bedrock it is possible to finalize the withdrawal only at the end of the dispute period (currently 7 days), after which a root is automatically considered valid. The length of this period is determined by the fact that proof of invalidity can be censored on Ethereum until its end. The success probability of this type of attack decreases exponentially as time increases:
$$
\mathbb{E}[\text{subtracted value}] = V * p^n
$$
where $n$ is the number of blocks in an interval, $V$ is the amount of funds that can be subtracted by publishing an invalid root, and $p$ is the probability of successfully performing a censorship attack in a given block. Suppose that this probability is $99\%$, that the value enclosed in the Rollup is one million Ether, and that the blocks in an interval are 1800 (6 hours of blocks with 12-second interval): the expected value is about 0.01391 Ether. The system is made secure by asking proposers of new roots to stake a much larger amount of Ether than the expected value.

Winzer et al. show how to carry out a censorship attack using a simple smart contract that ensures that certain areas of memory in the state do not change \cite{winzer2019temporary}. Modeling the attack as a Markov game, the paper shows that censoring is the dominant strategy for a rational block producer if they receive more compensation than including the transaction that changes memory. The $p$ value discussed just above can be viewed as the percentage of rational block producers in the network, where ``rational'' means that they follow short-term profit strategies but do not take into account possibly penalizing externalities, such as less trust in the blockchain that decreases its cryptocurrency price.

\vspace{0.35cm}

\begin{lstlisting}[caption={Example of a contract that incentivizes a censorship attack on Bedrock.},captionpos=b][language=Solidity]
function claimBribe(bytes memory storageProof) external {
    require(!claimed[block.number], "bribe already claimed");
    OutputProposal memory current = storageOracle.getStorage(L2_ORACLE, block.number, SLOT, storageProof);
    require(invalidOutputRoot == current.outputRoot, "attack failed");
    (bool sent, ) = block.coinbase.call{value: address(this).balance / period}("");
    require(sent, "failed to send ether");
}
\end{lstlisting}

The length of the dispute period must also take into account the fact that the invalidity proof is an interactive proof and therefore enough time must be provided for participants to interact and that any interaction may be censored. If the last move occurs at a time very close to the end of the dispute period, the cost of censoring is significantly less.

Although censoring is the dominant strategy, the likelihood of success decreases because censoring nodes are vulnerable to Denial of Service (DoS) attacks: an attacker can generate very complex transactions that end with the publication of an invalidity proof (causing the censoring attack to fail) at no cost, as no fees would be paid \cite{hess2016dos}. Because of the halting problem, there is no tool that can determine the outcome of a transaction without executing it. 

In extreme cases, a long dispute period allows coordination in the event of a successful censorship attack to organize a soft fork and exclude the attacking block producers. This defense is most effective using a PoS consensus algorithm that allows for a \textit{slash} on the stake of block proposers.

Another insidious attack, from which StarkNet and other Validity Rollups are immune, is to publish more state root proposals than disputants can verify. This can be avoided by using a frequency limit.

\subsection{Fast optimistic withdrawals}
Since the validity of an Optimistic Rollup can be verified at any time by anyone, a trusted oracle can be used to know on L1 whether the withdrawal can be finalized safely. This mechanism was first proposed by Maker \cite{macpherson2021fastbridge}: an oracle verifies the withdrawal, publishes the result on L1 on which an interest-bearing loan is assigned to the interested user, which is automatically closed at the end of 7 days, i.e., when the withdrawal can actually be finalized. This solution introduces an assumption of trust, but in the case of Maker it is minimized because the oracle operator is managed by the same organization that assumes the risk by providing the loan.

\section{Recursion: L3 and beyond}

At a high level, a validity proof can be seen as:
\begin{enumerate}
    \item The Prover generates the proof:
    $$P(A, x, y, w, T) = \pi$$
    where $A$ is a program, $x$ is the input, $y$ is the output, $w$ is the trace of the execution of program $A$ on input $x$ with time limit $T$. In the case of STARK, as seen in ~\ref{stark}, the time to generate $pi$ is $O(T \cdot \text{poly} \log(T))$.
    \item The Verifier verifies the proof:
    $$V(A,x,y,T,\pi) \in \{\text{TRUE}, \text{FALSE}\}$$
    In the case of STARK, the verification time is $O(\text{poly} \log(T))$.
\end{enumerate}

Since it is possible to generate proofs for arbitrary program computations and the verification of a proof is one of them, it is possible to recursively generate the proof of verification of a proof:

$$
    P(V, (A,x,y,T,\pi), \text{TRUE}, w_1, T_1)
$$

SHARP can aggregate hundreds of thousands of transactions into a single proof, where the bottleneck is determined by computational resources, such as memory. In addition, if you want to aggregate a high number of transactions you have to wait for the last transaction to arrive to start generating the proof. A recursive proof system is able to solve these problems by building a binary tree of proofs: transactions are provable in parallel, avoiding waiting for the last one and consuming excessive resources on a single demonstrator, and the time to obtain the final proof is strictly less than the standard method. 

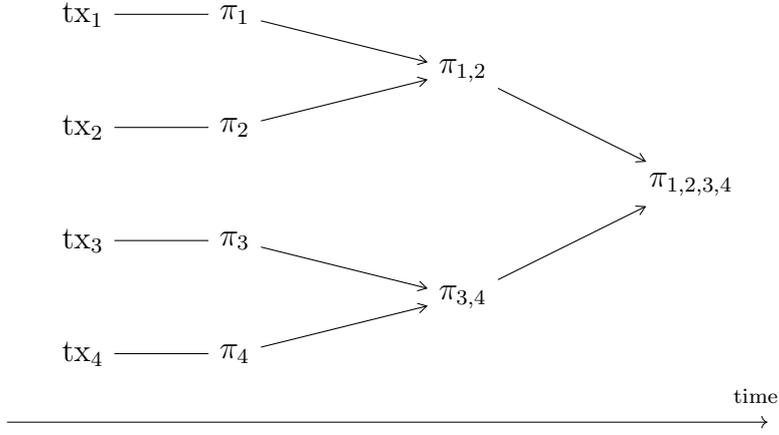
\vspace{0.35cm}

\begin{figure}[H]
\centering
\tikzstyle{level 1}=[level distance=30mm, sibling distance=30mm]
\tikzstyle{level 2}=[level distance=30mm, sibling distance=15mm]
\tikzstyle{level 3}=[level distance=20mm]
\begin{tikzpicture}[grow=left,<-,>=angle 60]
\begin{scope}[yshift=0]
  \node {$\pi_{1,2,3,4}$}
    child {node {$\pi_{1,2}$}
      child {node {$\pi_1$}
        child[-] {node{$\text{tx}_1$}}  
      }
      child {node{$\pi_2$}
        child[-] {node{$\text{tx}_2$}}  
      }
    }
    child {node {$\pi_{3,4}$}
      child {node{$\pi_3$}
        child[-] {node{$\text{tx}_3$}}  
      }
      child {node{$\pi_4$}
        child[-] {node{$\text{tx}_4$}}  
      }
    };
\end{scope}
\end{tikzpicture}

\begin{tikzpicture}
\draw[line width=0.1mm, ->] (0,0) -- (10,0) node[font=\scriptsize,above left=3pt and -8pt]{time};

\end{tikzpicture}
\end{figure}

Let $t$ be the time used to prove a transaction and $k$ the number of transactions, the standard proof system takes about $kt$ time to create an aggregate proof.
Asymptotically, the time to create a proof of transaction verification is less than the time of the original proofs:
$$
    O(T \cdot \text{poly} \log(T)) \supset O(2\cdot \text{poly}\log(T) \cdot \text{poly} \log(2 \cdot \text{poly} \log(T)))
$$
and this is true even in the specific instances of the problem until the limit imposed by the overhead \footnote{the recursive application of a logarithm summed to a constant converges} is reached. Assuming that the recursion stops before this limit, the proof time $t_i$ used at step $i$ is greater than that at step $i+1$:
$$
    t_1 > t_2 > t_3 > \dots > t_n
$$
from which follows that
$$
    n \cdot t_1 > t_1 + t_2 + t_3 + \cdots + t_n
$$

that is, $n \cdot t_1$ is an upper bound on the total proof time. The number of recursion steps is at most the height of the binary tree: given $k$ transactions it is approximately $\log(k)$. Since $t_1 = t$, for $k > 0$ and $t \geq 0$ it holds:
$$
    \log(k) \cdot t \leq kt
$$

Using this mechanism, the final Verifier needs to verify only recursive proofs instead of arbitrary programs, reducing complexity. 

By developing a SHARP Verifier in Cairo, one can leverage this mechanism for building an L3, which is an additional Rollup that relies on StarkNet. The latency of finalization on Ethereum is not significantly increased, and publishing data and proving proofs is less expensive as you interact with L2. The main advantage of an L3 is the ability to build Rollups for specific applications or with experimental technologies, without the need for a large number of transactions to amortize costs.

In contrast to Validity Rollups, building an Optimism Rollup on top of another Optimism Rollup does not lead to many advantages: the finalization period of an L3 adds up to that of L2, in the case of an Optimism Bedrock instance based on the main would become 14 days. Since function calls are published in a compressed version on Ethereum and any L3 would have to do the same on L2, there is no benefit to applying compression twice on strings.

\section{Transaction costs}

The cost of transactions on L2 is mostly determined by the interaction with L1 while the cost of what is performed off-chain is mostly negligible: for example, the price of gas on Optimism is stable on the order of a thousandth of a Gwei \cite{optimism2022gasprice}. StarkNet publishes on Ethereum in the form of a calldata every change in L2 storage and performs validity proof checks, while Optimism publishes in the form of a compressed calldata all the calldata of Rollup transactions and very rarely performs invalidity proofs \footnote{if the economic model works correctly and all participants are rational, one never has the need to publish them}. In both solutions the cost of computation is very cheap as it is executed entirely off-chain, but for StarkNet the storage used by transactions is the resource with the highest cost, while for Optimism the calldata is. 

As defined in the Ethereum Yellow Paper and in EIP-2028 \cite{akhunov2019eip2028}, the calldata cost on Ethereum for a byte representing zero is 4 gas and 16 per non-zero byte. The calculation of the gas cost of writing to storage on Ethereum depends on whether the cell has already been accessed (\verb|SLOAD|) or not, as shown below:

\renewcommand{\arraystretch}{1.4}
\begin{table}[H]
\begin{center}
\begin{tabular}{|l|c|c|}
\hline
\textbf{OPCODE}               & \textbf{with access} & \textbf{without access} \\ \hline
\verb|SSTORE| from zero to non-zero     & 22100                & 20000                  \\ \hline
\verb|SSTORE| from non-zero to non-zero & 5000                 & 2900                   \\ \hline
\verb|SSTORE| to byte zero &
  \begin{tabular}[c]{@{}c@{}}prev. value cost\\  + refund\end{tabular} &
  \begin{tabular}[c]{@{}c@{}}prev. value cost\\ + refund\end{tabular} \\ \hline
\verb|SSTORE| of a modified value  & 100                  & 100                    \\ \hline
\end{tabular}
\caption{\label{tab:SSTOREcost}Cost of modifying a storage cell. Calculating the refund is beyond the scope of this section.}
\end{center}
\end{table}

StarkNet publishes storage difference data in the format described in ~\ref{starknetDA}. The exact cost of the calldata represented differences depends on the particular values of the cells and the values that are written. Assuming no contract is published and 10 cells not previously accessed on StarkNet are modified in the manner described in Appendix ~\ref{appendiceCalldataStarknet} a cost of 9240 gas is calculated compared to 221000 if the storage had been modified on Ethereum, or about 4.18\% \footnote{the code used for the StarkNet estimate can be visited at the link \url{https://github.com/lucadonnoh/starknet-data-availability-cost}.}. If a cell is overwritten $n$ times between two data publications, the cost of each write will be $1/n$ compared to the cost of a single one since only the last change is published. The cost can be further minimized by compressing frequently used values. The cost of validity proof verification is divided among the transactions it refers to: for example, StarkNet block 4779 contains 200 transactions and its validity proof consumes 267830 gas, or 1339.15 gas for each.

Optimism publishes calldata of L2 transactions on L1 in the format described in ~\ref{opDA}. There is currently no network using the Bedrock version yet, but due to the equivalence with EVM, it is possible to test compression using the blocks of L1: assuming using the transactions in the Ethereum block height 13100000 (29904951 gas) to build a batch, compression with ZLIB reduces the size in bytes of the calldata by 51.69\% and the cost in gas by 79.74\% (from 1323384 to 1055548). The difference between the two ratios occurs because of the long sequences of zeros. The compression ratio can be further improved by compressing multiple batches together: using blocks 13100000 to 13100009 yields a compression in bytes of 51.02\% and in gas of 77.10\% \footnote{the code used for Optimism estimation can be visited at the link \url{https://github.com/lucadonnoh/optimism-data-availability-cost}.}. Assuming the price of gas on Optimism at 0.001 Gwei and the price of gas on Ethereum at 10 Gwei, the cost to process the 13100000 block on Optimism is about 3.5\% compared to if it were executed on Ethereum. 

\subsection{Optimizing the calldata: cache contract}

When designing a contract on Optimism one must keep in mind the fact that calldata is the most expensive resource. Presented below is a smart contract that implements an address cache that takes advantage of the fact that storage and execution are much less expensive resources along with a \verb|Friends| contract that demonstrates its use. The latter keeps track of the ``friends'' of an address that can be registered by calling the \verb|addFriend| function. If an address has already been used at least once, it can be added by calling the \verb|addFriendWithCache| function: the cache indices are 4-byte integers while the addresses are represented by 20 bytes, so there is an 80\% savings on the function argument. The same logic can be used for other data types such as integers or more generally bytes.

\begin{lstlisting}[caption={Address cache contract.},captionpos=b][language=Solidity]
contract AddressCache {
    mapping(address => uint32) public address2key;
    address[] public key2address;

    function cacheWrite(address _address) internal returns (uint32) {
        require(key2address.length < type(uint32).max, "AddressCache: cache is full");
        require(address2key[_address] == 0, "AddressCache: address already cached");
        // keys must start from 1 because 0 means "not found"
        uint32 key = uint32(key2address.length + 1);
        address2key[_address] = key;
        key2address.push(_address);
        return key;
    }

    function cacheRead(uint32 _key) public view returns (address) {
        require(_key <= key2address.length && _key > 0, "AddressCache: key not found");
        return key2address[_key - 1];
    }
}
\end{lstlisting}

\begin{lstlisting}[caption={Example of a contract that inherits the address cache.},captionpos=b][language=Solidity]
contract Friends is AddressCache {
    mapping(address => address[]) public friends;

    function addFriend(address _friend) public {
        friends[msg.sender].push(_friend);
        cacheWrite(_friend);
    }

    function addFriendWithCache(uint32 _friendKey) public {
        friends[msg.sender].push(cacheRead(_friendKey));
    }

    function getFriends() public view returns (address[] memory) {
        return friends[msg.sender];
    }
}
\end{lstlisting}

The contract supports in cache about 4 billion ($2^{32}$) addresses, and adding one byte gives about 1 trillion ($2^{40}$).

\subsection{Optimizing storage: Bloom's filters}

On StarkNet there are several techniques for minimizing storage usage. If it is not necessary to guarantee the availability of the original data then it is sufficient to save the on-chain hash: this is the usual mechanism used to save data for an ERC-721 (NFT) \cite{entriken2018eip721}, i.e., an IPFS link that resolves the hash of the data if available. For data that is stored multiple times, it is possible to use a look-up table similar to the caching system introduced for Optimism but requiring that all values to be saved at least once. For some applications this can be avoided by using a Bloom filter \cite{bloom1970space} \cite{christensen2010new} \cite{agarwal2006approximating}, i.e., a probabilistic data structure that allows one to know with certainty whether an element does not belong to a set but admits a small but non-negligible probability of false positives.

A Bloom filter is initialized as an array of $m$ bits at zero. To add an element $k$ hash functions with a uniform random distribution are used, each one mapping to a bit of the array that is set to 1. To check whether an element belongs to the set we run the $k$ hash functions and verify that the $k$ bits are set to 1. In a simple Bloom's filter there is no way to distinguish whether an element actually belongs to the set or is a false positive, a probability that grows as the number of entries increases. After inserting $n$ elements:
$$\mathbb{P}[\text{false positive}] = \left(1-\left[1-\frac{1}{m}\right]^{kn}\right)^k \approx \left(1-e^{-kn/m}\right)^k$$

assuming independence of the probability of each bit set. If $n$ elements (of arbitrary size!) are expected to be included and the probability of a false positive tolerated is $p$, the size of the array can be calculated as:
$$m = -\frac{n \ln p}{(\ln 2)^2}$$

While the optimal number of hash functions is:
$$k = \frac{m}{n} \ln 2$$

If we assume to insert 1000 elements with a tolerance of 1\% the size of the array is 9585 bits with $k = 6$, while for a tolerance of 0.1\% it becomes 14377 with $k=9$. If a million elements are expected to be inserted, the size of the array becomes about 1170 kB for 1\% and 1775 kB for 0.1\%, with the same values of $k$, since it depends only on $p$ \cite{starobinski2003efficient}.

In a game where players must not be assigned to an opponent they have already challenged, instead of saving in storage for each player the list of past opponents one can use a Bloom filter. The risk of not challenging some players is often acceptable, and the filter can be reset periodically. An implementation of a Bloom filter in Cairo was developed by Sam Barnes \cite{barnes2022bloom}.


\section{Ethereum compatibility}

The main advantage of being compatible with EVM and Ethereum is the reuse of all tools. Ethereum smart contracts can be published on Optimism without any modifications and new audits: in fact, major applications such as Synthetix, Uniswap, Aave, and Curve are already active. Wallets remain compatible, development and static analysis tools, general analysis tools, indexing tools and oracles. Ethereum and Solidity have a long history of well-studied vulnerabilities, such as reentrancy attacks, overflows and underflows, flash loans, and oracle manipulations that have resulted in attacks where tens or thousands of millions of dollars have been stolen \cite{cream2021hack} \cite{pickle2020hack}. Because of this Optimism was able to capture a large amount of value in a short time.

Choosing to adopt a different virtual machine implies having to rebuild an entire ecosystem but with the advantage of having a greater implementation freedom. StarkNet natively implements \textit{account abstraction}, which is a mechanism whereby each account is a smart contract that can implement arbitrary logic as long as it complies with an interface (hence the term \textit{abstraction}): this allows for different digital signature schemes, the ability to change the private key using the same address, or use a multisig. The Ethereum community proposed the introduction of this mechanism with EIP-2938 in 2020 but the proposal has remained stale for more than a year as other updates have been given more priority \cite{buterin2020eip2938}.

Another important benefit gained from compatibility is the reuse of existing clients: Optimism uses a version of geth for its own node with only ~800 lines of difference, which has been developed, tested, and maintained since 2014. Having a robust client is crucial as it defines what is accepted as valid or not in the network. A bug in the implementation of the invalidity proof system could cause an incorrect invalidity proof to be accepted as correct or an invalidity proof for an invalid block to be accepted as incorrect, compromising the system. 
The likelihood of this type of attack can be limited with a wider diversity of clients: Optimism can reuse in addition to geth the other Ethereum clients already maintained, and development of another Erigon-based client is already underway. 
In 2016 a problem in the memory management of geth was exploited for a DoS attack and the first line of defense was to recommend the use of Parity, the second most used client at the time.
StarkNet has the same problem with validity proofs, but the clients have to be written from scratch and the proof system is much more complex and consequently it is also much more complex to ensure correctness. Currently StarkNet has only one client and no others are being developed.

\section{License}

StarkNet has often been the focus of attention because of its restrictive license. The Cairo language and toolchain adopt the Cairo Toolchain License \cite{cairo2022license}: it allows the use of Cairo for writing and compiling Cairo programs and other non-commercial uses such as academic and scientific research. The toolchain is modifiable only for fixing bugs and not for introducing new features, and copying and distribution of the code (which is publicly available) is not permitted.

The Prover code, which is currently closed source, will be released under a StarkWare Polaris Prover License \cite{prover2022license} that allows commercial use only to generate proofs that are sent to a Polaris Verifier, which is a StarkWare-approved Verifier. The list of approved Polaris Verifiers can only be extended, so an approved Verifier cannot be revoked. This effectively prevents the creation of forks independent of StarkWare.

Optimism, in contrast, is completely open source and uses an MIT license for all the tools used. This has led to the creation of two forks, Metis Andromeda \cite{metis2022fork} and Boba Network \cite{boba2022fork}, but these have not compromised the success of the original version. This has already been seen with Bitcoin and especially Ethereum, which feature numerous forks that have only legitimized the technology and led more developers to verify and expand the original code. Currently the EVM is the most widely used virtual machine for blockchain and Solidity the most widely used language for writing smart contracts, and Optimism aims in the same way to become a de facto standard for building Rollups.

\chapter{Conclusion}

Rollups are the most promising solution available today to solve the scalability problem in decentralized blockchains, ushering in the era of modular blockchains as opposed to monolithic blockchains.

The choice of developing an Optimistic Rollup or a Validity Rollup is mainly shown as a trade-off between complexity and agility. StarkNet has numerous advantages such as potentially instantaneous withdrawal speed, structural inability to have invalid state transitions, recursion and lower transaction cost at the expense of a longer development period and incompatibility with EVM, where instead Optimism leveraged the network economy to quickly gain a major share of the market.

Optimism Bedrock, however, possesses a modular design that allows it to become a Validity Rollup in the future: Cannon currently uses \verb|minigeth| compiled to MIPS for invalidity proof contention, but the same architecture can be used to obtain a circuit and produce validity proofs. Compiling a complex machine such as the EVM for a microarchitecture results in a simpler circuit that does not need to be modified and re-verified in case of upgrades. RISC Zero is a verifiable microarchitecture with STARK proofs already in development based on RISC-V that can be used for this purpose as an alternative to MIPS \cite{risc2022zero}.

One aspect that should not be underestimated is the complexity in understanding how the technology works. A strength of traditional blockchains is being able to verify the state of the blockchain without trusting any third party entity, but in the case of StarkNet even if you use your own node you have to trust the implementation if you are unable to verify the various components based on cryptography and advanced mathematics. This may initially create friction for the adoption of the technology, but as the tools and adoption of integrity proofs advance even outside the blockchain field this problem will be hopefully solved.

\chapter*{Appendix}

\section*{StarkNet calldata value}
\label{appendiceCalldataStarknet}

[0, 1, 78012987367078498244736967587441276376014206154405857948822581408104104410721, 10, 49437887447255105617199385887980129590299043410906399897274339686664380574960, 81613196144862953930755284412013485753825942725888221915012079651792110103808, 77869845672245121662237546936898195077685970774400528945790634750486399986245, 85558286294651018119282355933772523799565789757486469436870233741200601720903, 90745439112799995280673958963319809841091902573630903294655608952911237510638, 49, 72063704605688213715872376071514311689316615270384662374827175421482880125180, 39047936296155467891523306114750972410898988810559128988743926746334839389254, 89821206671539319279995197695429264123175493398319804842575199728181115252599, 99, 47475753046911164737671950579172075423187336110653749106497219281656544366808, 29, 30594499811872827545153257993174147177746163003834628645239607985359843108205, 16, 21230045744089919195261861661020416944848194956527998680880953029066897219408, 27941555059559098141567348626988165098886309475575494710999032236178114317593, 83549733318410479614820445166391282086750526240790917555062354500545869380230, 17, 70199979574190103393325973797566928885460655906709293378713100338207628138006, 3702205553337436218648230511058213631110329670271146471049479018502731771592]

i.e.:
\begin{itemize}
    \item Cells dedicated to contract deployments: 0
    \item Number of contracts with modified cells: 1
    \item Contract address
    \item Number of modified cells: 10
    \item List of modified cells (key, value), including 5 with small values ($\leq 100$) and 5 with high values (up to $2^{256}-1$).
    
\end{itemize}

\printbibliography

\end{document}